\documentclass[twocolumn]{aastex631}

\begin{document}

\title{Thermal electrons in the radio afterglow of relativistic tidal disruption event ZTF22aaajecp/AT2022cmc}

\author[0000-0002-0786-7307]{Lauren Rhodes}
\affiliation{Trottier Space Institute at McGill, 3550 Rue University, Montreal, Quebec H3A 2A7, Canada}
\affiliation{Department of Physics, McGill University, 3600 Rue University, Montreal, Quebec H3A 2T8, Canada}
\author[0000-0001-8405-2649]{Ben Margalit}
\affiliation{School of Physics and Astronomy, University of Minnesota, Minneapolis, MN 55455, USA}
\author[0000-0002-7735-5796]{Joe S. Bright}
\affiliation{Astrophysics, Department of Physics, The University of Oxford, Keble Road, Oxford, OX1 3RH, UK}
\affiliation{Breakthrough Listen, Astrophysics, Department of Physics, The University of Oxford, Keble Road, Oxford, OX1 3RH, UK}
\author[0009-0008-6396-0849]{Hannah Dykaar}
\affiliation{Trottier Space Institute at McGill, 3550 Rue University, Montreal, Quebec H3A 2A7, Canada}
\affiliation{Department of Physics, McGill University, 3600 Rue University, Montreal, Quebec H3A 2T8, Canada}
\author{Rob Fender}
\affiliation{Astrophysics, Department of Physics, The University of Oxford, Keble Road, Oxford, OX1 3RH, UK}
\affiliation{Department of Astronomy, University of Cape Town, Private Bag X3, Rondebosch 7701, South Africa}
\author[0000-0003-3189-9998]{David A. Green}
\affiliation{Astrophysics Group, Cavendish Laboratory, 19 J.J. Thomson Avenue, Cambridge CB3 0HE, UK}
\author[0000-0001-6803-2138]{Daryl Haggard}
\affiliation{Trottier Space Institute at McGill, 3550 Rue University, Montreal, Quebec H3A 2A7, Canada}
\affiliation{Department of Physics, McGill University, 3600 Rue University, Montreal, Quebec H3A 2T8, Canada}
\author[0000-0002-5936-1156]{Assaf Horesh}
\affiliation{Racah Institute of Physics, The Hebrew University of Jerusalem, Jerusalem 91904, Israel}
\author[0000-0001-9149-6707]{Alexander J. van der Horst}
\affiliation{Department of Physics, George Washington University, 725 21st St NW, Washington, DC 20052, USA }
\author[0000-0003-0764-0687]{Andrew Hughes}
\affiliation{Astrophysics, Department of Physics, The University of Oxford, Keble Road, Oxford, OX1 3RH, UK}
\author[0000-0002-2557-5180]{Kunal Mooley}
\affiliation{Indian Institute Of Technology Kanpur, Kanpur, Uttar Pradesh 208016, India}
\affiliation{Caltech, 1200 E. California Blvd. MC 249-17, Pasadena, CA 91125, USA}
\author[0000-0003-0466-3779]{Itai Sfaradi}
\affiliation{Department of Astronomy, University of California, Berkeley, CA 94720-3411, USA}
\author{David Titterington}
\affiliation{Astrophysics Group, Cavendish Laboratory, 19 J.J. Thomson Avenue, Cambridge CB3 0HE, UK}
\author[0000-0001-7361-0246]{David Williams-Baldwin}
\affiliation{Jodrell Bank Centre for Astrophysics, School of Physics and Astronomy, The University of Manchester, Manchester, M13 9PL, UK}



\begin{abstract}
A tidal disruption event (TDE) occurs when a star travels too close to a supermassive black hole. In some cases, accretion of the disrupted material onto the black hole launches a relativistic jet. In this paper, we present a long term observing campaign to study the radio and sub-millimeter emission associated with the fifth jetted/relativistic TDE: AT2022cmc. Our campaign reveals a long lived counterpart. We fit three different models to our data: a non-thermal jet, a spherical outflow consisting of both thermal and non-thermal electrons, and a jet with thermal and non-thermal electrons. We find that the data is best described by a relativistic spherical outflow propagating into an environment with a density profile following $R^{-1.8}$. Comparison of AT2022cmc to other TDEs finds agreement in the density profile of the environment but also that AT2022cmc is twice as energetic as the other well-studied relativistic TDE \textit{Swift} J1644. Our observations of AT2022cmc allow a thermal electron population to be inferred for the first time in a jetted transient providing, new insights into the microphysics of relativistic transients jets.
\end{abstract}



\section{Introduction} \label{sec:intro}

In a scenario where a star travels too close to a supermassive black hole (SMBH) in the center of a galaxy, the gravitational influence of the SMBH overcomes the binding energy keeping that star together and creates a tidal disruption event \citep[TDE, ][]{1988Natur.333..523R}. Roughly half of the disrupted stellar material is lost and the rest remains gravitationally bound to the system. As the material accretes onto the central black hole, it sometimes generates and launches an outflow, which may be a relativistic jet \citep{2011MNRAS.416.2102G}. There have been five TDEs to date interpreted as having a relativistic jet \citep[\textit{Swift} J1112.2-8238, \textit{Swift} J2058.4+0516, \textit{Swift} J164449.3+573451, Arp 299-B AT1][]{2011Natur.476..425Z, 2012ApJ...753...77C, doi:10.1126/science.aao4669, 2015MNRAS.452.4297B}, the most recent of which was ZTF22aaajecp/AT2022cmc (hereafter AT2022cmc), the subject of this study.

%\subsection{AT2022cmc} \label{sec:cmc}

AT2022cmc was a tidal disruption event discovered by the Zwicky Transient Facility on 2022 February 11 10:42 UT \citep[MJD 59621.4458, $T_{0}$, ][]{2022TNSAN..38....1A}. At the time of writing, of the five confirmed relativistic TDEs, AT2022cmc was the most distant at $z = 1.193$ \citep[assuming $H_0$ = 70\,km\,s$^{-1}$\,Mpc$^{-1}$ and $\Omega_{\rm{M}}$ = 0.3, ][]{2022GCN.31602....1T}. Subsequently, AT2022cmc has been observed across the electromagnetic spectrum \citep{2022Natur.612..430A,2023MNRAS.521..389R,2023NatAs...7...88P,2024ApJ...974..149E,2024ApJ...965...39Y}. The X-ray counterpart to AT2022cmc was highly variable with large flares lasting thousands of seconds \citep{2023NatAs...7...88P} until around 400\,days post-discovery when the X-ray flux dropped by at least an order of magnitude \citep{2024ApJ...974..149E}. Radio observations covering the first 100\,days post-discovery revealed an optically thick ($\gamma \geqslant 2$, $F_{\nu}\propto\nu^{\gamma}$), evolving counterpart \citep{2022Natur.612..430A, 2023MNRAS.521..389R}. \citet{2023MNRAS.521..389R} showed that the radio counterpart had a bulk Lorentz factor of at least 8 by studying the variability observed at 15.5\,GHz and \citet{2023NatAs...7...88P} found an even higher Lorentz factor of $\sim$90 by applying a blazar model to the multi-wavelength data. These very high Lorentz factors led investigators to conclude that there had to be a jet in the system, where the drop in X-ray flux indicated that the jet was switching off.

To explain the behavior in different wave bands, a number of different scenarios have been invoked \citep[e.g.][]{2022NatAs.tmp..252P,2024ApJ...965...39Y} but in all scenarios non-thermal synchrotron emission drives the modelled emission. For example, \citet{2022Natur.612..430A} explained the evolving radio counterpart with synchrotron emission as is observed in gamma-ray burst (GRB) afterglows \citep{2002ApJ...568..820G}.In GRB afterglow models, a highly relativistic, decelerating jet sweeps up electrons in the circumburst environment accelerating them across a shock front into a power law energy distribution ($N(E)\,dE \propto E^{-p} \,dE$) and then cools, emitting synchrotron emission which is brightest in the radio band \citep[][]{1976PhFl...19.1130B, 1997ApJ...476..232M, 1997ApJ...489L..37S, 1998ApJ...497L..17S}. It is often, but not always, assumed that all the electrons are part of this non-thermal distribution \citep{2005ApJ...627..861E}. The electrons cool and emit synchrotron emission that has a well-described spectrum following a number of power laws where the peak of the spectrum is optically thin and allows observers to track the electron energy distribution down to the lowest electron Lorentz factors \citep{2002ApJ...568..820G, 2023MNRAS.518.1522D}. These models are valid from the regime of ultra-relativistic jets through to a non-relativistic phase and was also used to explain \textit{Swift} J1644$+$57, the most well-studied relativistic TDE to date. 

Models have been developed that consider a thermal electron population in addition to the non-thermal population \citep{2009MNRAS.400..330G, 2017ApJ...845..150R, 2017ApJ...835..248W, 2018MNRAS.480.4060W, 2022ApJ...924...40W, 2021ApJ...923L..14M, 2024ApJ...977..134M}. Such models may be particularly important for mildly relativistic outflows where the shock velocity is $0.2 \lesssim (\Gamma\beta)_{\rm{sh}} \lesssim 2$ ($(\Gamma\beta)_{\rm{sh}}$ is the shock proper velocity), such as those produced in Fast Blue Optical Transients \citep{2022ApJ...932..116H}. Recent theoretical work has also explored this in the context of ultra-relativistic GRB afterglows, however, thermal electrons are typically ignored when modelling GRB observations \citep{2009MNRAS.400..330G, 2017ApJ...845..150R, 2017ApJ...835..248W, 2018MNRAS.480.4060W, 2022ApJ...924...40W}. The inclusion of synchrotron-emitting thermal electrons has not been explored in the modelling of TDEs.

In this paper, we present the results of a long term radio and sub-millimeter monitoring campaign on the most recently discovered relativistic TDE, AT2022cmc. In Section \ref{sec:obs}, we present the details of our monitoring campaign along with the data reduction methods used; in Sections \ref{sec:results} and \ref{sec:Modelling}, we present the results of our campaign and the different models used to explain our findings. In Section \ref{sec:discussion}, we discuss the implications of these fits and contextualise them within both the TDE literature and synchrotron transients as a whole. Finally, we present our conclusions in Section \ref{sec:conc}.

\section{Observations} \label{sec:obs} 

Here, we summarise the observations and the data reduction methods used to study AT2022cmc. All the resulting flux densities and 3$\sigma$ upper limits are given in Table \ref{tab:data}. In our analysis of this source, we also include data published by \citet{2022Natur.612..430A}.

\subsection{AMI--LA}

The Arcminute Microkelvin Imager -- Large Array (AMI--LA) is an eight-dish interferometer based in Cambridge, UK \citep{amila, 2018MNRAS.475.5677H}. It observes at a central frequency of 15.5\,GHz with a bandwidth of 5\,GHz, achieving an angular resolution of ${\sim}\,$30\,arcsec. Whilst the first 100\,days of observations have already been reported in \citet{2023MNRAS.521..389R}, we continued observing with AMI--LA until October 2024. AMI--LA data is reduced using a custom software package: \textsc{reduce\_dc} \citep{2013MNRAS.429.3330P}. Given AT2022cmc's proximity on the sky to 3C286, all the flux scaling, bandpass and complex gain calibration is conducted using 3C286. Further flagging, cleaning and deconvolution is done in \textsc{casa} using the tasks \textit{rflag}, \textit{tfcrop} and \textit{tclean} \citep{casanew}.

For observations until the end of January 2024, each observation was about four hours long and the starting times, dates, and flux densities are quoted in Table \ref{tab:data}. From February 2024 onwards, as a result of the low signal to noise and non-detections, we concatenated multiple epochs. This was done for three sets of observations, the first included all of the observations from March 2024, the second included those in July and August 2024, and the final set consisted of all epochs in April 2025. In the three stacked observations we obtain flux densities of 0.15$\pm$0.01, 0.13$\pm$0.01 and 0.11$\pm$0.3\,mJy, respectively. These three epochs are also included in Table \ref{tab:data} and Figure \ref{fig:lc}.

\subsection{e-MERLIN}

The \textit{enhanced} Multi-Element Remotely Linked Interferometer Network (\textit{e}-MERLIN) is a UK-based, radio interferometer consisting of seven dishes. With a maximum baseline of 217\,km, whilst observing at 5\,GHz (C-band, and bandwidth of 0.512\,GHz), \textit{e}-MERLIN can achieve an angular resolution of $0\farcs05$. We observed AT2022cmc with \textit{e}-MERLIN at C-band between February 2022 and August 2024 for a total of nine epochs (programs RRT13002, CY16004 and CY18002, PI: L. Rhodes).

All observations were reduced using the \textit{e}-MERLIN pipeline within \textsc{casa} \citep{CASA, 2021ascl.soft09006M}. The pipeline performs preliminary flagging for radio frequency interference (RFI) and known observatory issues. Flux density scaling is performed using 3C286 followed by bandpass calibration and complex gain calibration, using OQ~208 and J1905+1943, respectively. Further flagging of the target field is conducted. We performed interactive cleaning and deconvolution using the \textsc{casa} task \textit{tclean}.

In the first two epochs, we did detect any radio emission at the coordinates of AT2022cmc. From epoch 3 onwards, we consistently detect an unresolved radio source at a position consistent with those reported in the literature.

\subsection{MeerKAT}

MeerKAT is a 64-dish interferometer based in the Karoo Desert, South Africa. We obtained time on MeerKAT through two open time proposals (MKT-23101 and MKT-24207, PI: L. Rhodes) to observe at both L- (1.28\,GHz) and S4-band (3.01\,GHz), with bandwidths of 0.875 and 0.856\,GHz, respectively. In this work, we published all observations taken to date, from April 2022 until October 2024. We plan to continue monitoring this source in future observing terms. 

All MeerKAT observations were processed using \texttt{OxKAT}, a set of Python scripts specifically designed to reduce MeerKAT data \citep{oxkat}. Each observation is first averaged down to 1024 channels. The calibrator fields are flagged for RFI and then solved for amplitude and gain calibration solutions. The solutions are applied to the target fields, and then flagging and calibration are performed in \textsc{casa} and \textsc{tricolour}, respectively \citep{CASA, 2022ASPC..532..541H}. We also perform a round of phase-only self-calibration using \textsc{cubical}\footnote{https://github.com/ratt-ru/CubiCal} before imaging the field with \textsc{wsclean} \citep{offringa-wsclean-2014}.

\subsection{NOEMA}

The NOrthern Extended Millimetre Array (NOEMA), in the French Alps, monitored AT2022cmc through programs S22BT and W22CZ (PI: L. Rhodes) between June 2022 and April 2023. Observations were made in the 3\,mm band. The data was split into two sub-bands (86.25 and 101.75\,GHz) each with a bandwidth of 7.74\,GHz. 

Calibration was performed with \textsc{clic}, part of the \textsc{gildas}\footnote{\url{https://www.iram.fr/IRAMFR/GILDAS}} package. For each epoch, one of the sources 3C273, 3C345, J1310$+$323 or J2200$+$420 were used for bandpass calibration. J1328$+$307, J1302$-$102 or J1310$+$323 were then used for phase and amplitude calibration. The flux densities and their errors were derived from point-source UV-plane fits to the calibrated interferometric visibilities. Given the high signal-to-noise of the detections, we were able to measure the flux density in both sub-bands.

\section{Observational Summary} \label{sec:results}

\begin{figure*}
    \centering
    \includegraphics[width = 0.8\textwidth]{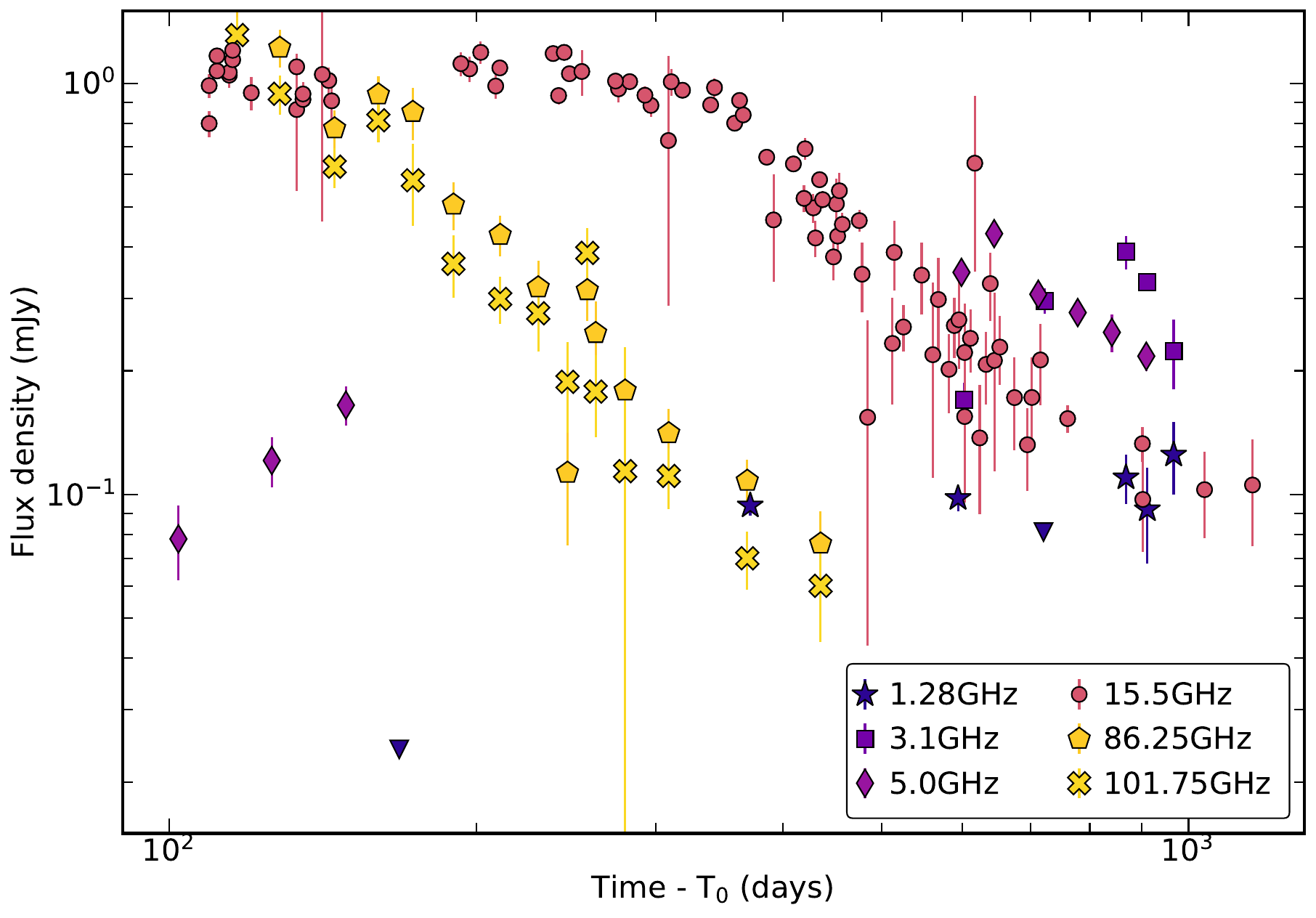}
    \caption{The radio and sub-mm light curve for AT2022cmc between 100 and $\sim$1000\,days post-discovery. At the highest frequencies, 86.26 and 101.75, the light curve follows a power law decay. At lower frequencies (at 15.5\,GHz and below) the light curves are best described by a broken power law. We quantify the rise and decay rates at each frequency in Section \ref{sec:results}.}
    \label{fig:lc}
\end{figure*}

\begin{figure*}
    \centering
    \includegraphics[width=0.8\textwidth]{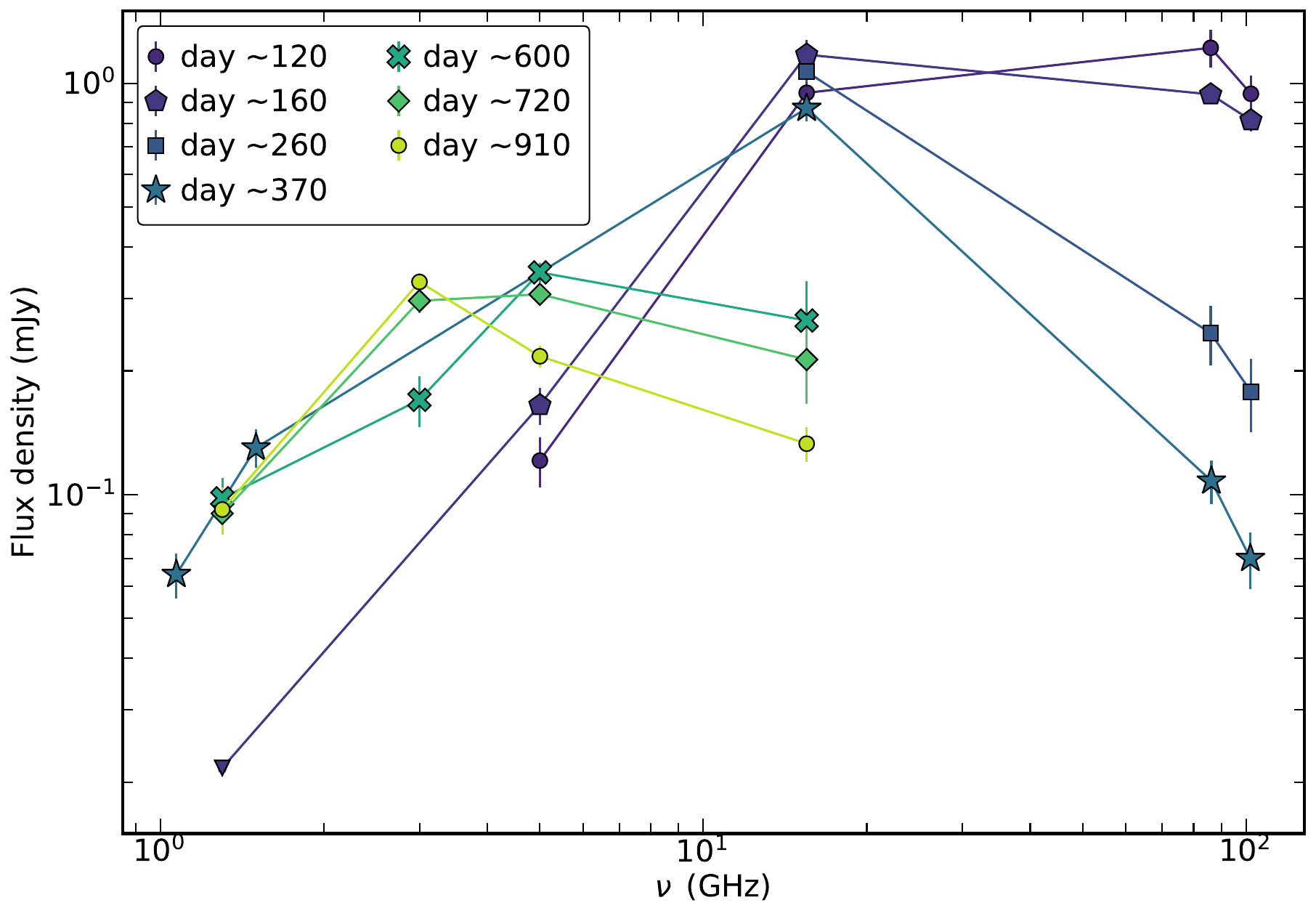}
    \caption{Spectral energy distributions (SEDs) constructed using data in Figure \ref{fig:lc} at 120, 160, 260, 270, 600 and 720\,days post-burst. There is a clear shift of the spectral peak over the duration of our campaign from above 15.5\,GHz ($<300$\,days) to around 3\,GHz ($\sim900$\,days).}
    \label{fig:seds}
\end{figure*}

Here, we consider the basic evolution of the radio and sub-mm counterpart by examining the evolution of the SED. Figures \ref{fig:lc} and \ref{fig:seds} display the data presented in Section \ref{sec:obs} as light curves and SEDs, respectively. To describe the behavior of AT2022cmc, we use the convention of $F_{\nu} \propto t^{\alpha}\nu^{\gamma}$ from 100\,days post-discovery. The light curves show that at the highest frequencies, 86.25 and 101.75\,GHz (pentagons and crosses, respectively), we observe the flux density decaying with a power law slope of $t^{-2.4\pm0.1}$. Moving to lower frequencies, in agreement with \citet{2022Natur.612..430A}, a break is visible in each band. At 15.5\,GHz, the peak in the light curve occurs at 300$\pm$8\,days, and the full 15.5\,GHz light curve is well described by a broken power law with slope indices of $t^{0.33\pm0.02}$ and $t^{-2.4\pm0.1}$. There is also evidence of a tentative flattening after 1000\,days.

Between 1 and 5\,GHz inclusive, we obtained fewer observations and we observe more dramatic changes in behavior. At both 3.1 and 5\,GHz (squares and diamonds, respectively), the light curve appears to have a peak at $\sim$800 and 650$\pm$40 days (we do not fit a broken power law to the 3.1\,GHz data due to the low number of data points). At 5\,GHz, over the first 200\,days there is variability which we cannot attribute to calibration errors. The temporal slopes follow $0.63\pm0.07$ and $-1.7\pm0.6$. At 1.28\,GHz, there is a rapid rise ($t^{1.6\pm0.7}$)\footnote{We note that one of the two data points used to calculate the rise is an upper limits and the uncertainty is calculated using the second equation in section 2.2 from \citet{2018MNRAS.473.4122E}} followed by a plateau ($t^{0.15\pm0.07}$).

In Figure \ref{fig:seds}, we have constructed radio and sub-mm spectral energy distributions (SEDs) at 7 different epochs (120, 160, 270, 370, 600, 720, 910\,days) to demonstrate the spectral evolution of the radio counterpart. The peak of the SED moves from higher to lower frequencies as time progresses: between 100-200\,days the peak sits between 15.5 and 100\,GHz but by 900\,days it is around 3\,GHz. We find that the low frequency branch of the SED does not change with time. A joint fit to all the SEDs gives $\gamma = 1.94\pm0.07$ for the low frequency branch. Conversely, the high frequency branch becomes shallower with time. The steepest spectral index we measure is $\gamma= -3\pm1$ at around 370\,days post-discovery between 86.25 and 101.75\,GHz but a joint fit to all the 86.25 and 101.75\,GHz (spanning 116 -- 436\,days post-discovery) data finds a shallower spectral index of $\gamma = -1.4\pm0.3$. By 900\,days we measure a spectral index of $\gamma = -0.4\pm0.1$ between 5 and 15.5\,GHz. There is a $3\sigma$ change in the high frequency spectral index over the course of our observing campaign.

\section{Modelling}\label{sec:Modelling}

In this Section, we consider the wealth of radio and sub-mm observations targeting AT2022cmc and interpret them within three different frameworks. We include the observations presented in Section \ref{sec:obs} in addition to those published in \citet{2022Natur.612..430A} and \citet{2023MNRAS.521..389R}.

\subsection{Non-thermal relativistic jet}\label{subsec:non_thermal_jet}

\citet{2023MNRAS.521..389R} showed that the early-time ($\lesssim100$ days) radio emission comes from an emitting region with a substantial bulk Lorentz factor ($\gtrsim$8). Hence, we first attempt to explain the radio emission using a gamma-ray burst (GRB) afterglow model. We compare the data at 15.5, 86.25 and 101.75\,GHz (where we have the best temporal coverage) to analytical results from GRB afterglow models \citep{2002ApJ...568..820G}.

At early times, a GRB jet is so relativistic that the only area visible to the observer is that within an opening angle of 1/$\Gamma$ (where $\Gamma$ is the bulk Lorentz factor). A jet break is an achromatic light curve signature observed in GRB afterglows when the jet has decelerated sufficiently such that the whole shock front is visible to the observer. Both the 86.25 and 101.75 light curves follow $t^{-2.4\pm0.1}$, which is consistent with a post-jet break light curve \citep[$t^{-p}$ where $p$ is the electron energy spectral index][]{1999ApJ...519L..17S}. The 15.5\,GHz (AMI--LA) decay also follows $t^{-2.4\pm0.1}$, consistent with a jet break. The 15.5\,GHz light curve rise which follows $t^{0.33\pm0.02}$ is consistent with a GRB afterglow where the observing frequency is below the characteristic electron frequency ($\nu_{\rm{m}}$) and above the self-absorption frequency \citep[$\nu_{\rm{sa}}$, ][]{2014MNRAS.444.3151V} in a pre-jet break regime. In order to create a self-consistent picture, a jet break would have to occur around the same time as $\nu_{\rm{m}}$ passing through the 15.5\,GHz observing band as only one light curve break is observed. Without the jet break, i.e. the peak is only a result of $\nu_{\rm{m}}$ moving through the band, we would expect the decay slope to follow $t^{-1.22\pm0.09}$. The two events occurring at the same time is not impossible but would require some fine tuning.

This scenario also presents problems when we use closure relations from e.g. \citet{2002ApJ...568..820G} to compare the value of $p$ from the high frequency (15.5, 86.25 and 101.75\,GHz) light curves ($2.4\pm0.1$) and the high frequency spectral index ($-1.4\pm0.3$). The light curves predict $p = 2.4\pm0.1$ \textit{for a post-jet break decay ($t^{-p}$)} but the spectral index predicts $p = 3.8\pm0.6$ ($\nu^{(1-p)/2}$). The two differ at greater than $2\sigma$. A spectral index of $-0.7$ is required to get $p = 2.4$ which we do not measure until at least 700\,days, significantly later than the time of our final NOEMA observation 370\,days post-discovery. We note that a similar issue was identified in \citet{2023MNRAS.522.4028M}, where the early optical spectral index was identified \citet[$\gamma = -1.32\pm0.18$][]{2022Natur.612..430A} and attributed to a fast cooling synchrotron scenario. Given how long we measured a steep spectral index, we find a fast cooling scenario unlikely and rule out a GRB-like scenario. 

The fine-tuning required to reproduce the turnover in the 15.5\,GHz light curve, combined with the discrepancy between the high-frequency decay and the spectral index, makes the GRB afterglow-like scenario difficult to reconcile with the observed radio emission from AT2022cmc. Therefore, we rule out this interpretation.

\subsection{Spherical outflow with thermal and non-thermal emitting particles}\label{subsec:thermal_non_thermal_sph}

Motivated by the steep spectral index at high frequencies, we consider the possible presence of a thermal electron population. We use the spherical outflow model presented in \citet{2024ApJ...977..134M}, an extension of \citet{2021ApJ...923L..14M}, that considers any shock velocity and deceleration profile (we consider a power-law profile: $(\Gamma\beta)_{\rm sh} \propto t^{-\alpha_{(\Gamma\beta)_{\rm sh}}}$), ranging from a Newtonian constant velocity scenario to an ultra-relativistic Blandford--McKee solution \citep{1976PhFl...19.1130B,1998ApJ...499..810C}. The outflow propagates into an environment parameterized by $n \propto r^{-k}$ allowing for a change in the normalization and density profile of the circumnuclear environment. Some fraction of the post-shock energy is in the thermal electron population ($\epsilon_{\rm{T}}$), whose temperature is set by the velocity of the shock, and some fraction is carried by a non-thermal population ($\epsilon_{\rm{e}}$) as described in Section \ref{subsec:non_thermal_jet} \citep{1998ApJ...499..810C}. The model calculates the emergent synchrotron luminosity considering both thermal and non-thermal electron populations, synchrotron self-absorption, and synchrotron cooling. 
This analytic model employs an effective line-of-sight approximation, analogous to \cite{1998ApJ...497L..17S}. A more detailed treatment that accurately integrates over emission from different regions of the shock (analogous to \citealt{1999ApJ...513..679G,1999ApJ...527..236G}) can only be performed numerically, and is the subject of forthcoming work (Ferguson \& Margalit in prep.).
The overall shape of the SED and the relative contributions of the different electron populations is a strong function of the shock velocity \citep{2024ApJ...977..134M}. The geometry of the source is assumed to be spherical. We note that this scenario is also applicable to a relativistic jet pre-jet break where the edges of the jet are not visible to the observer.

\begin{figure}
    \centering
    \begin{subfigure}%[]{}
        \centering
        \includegraphics[width = \columnwidth]{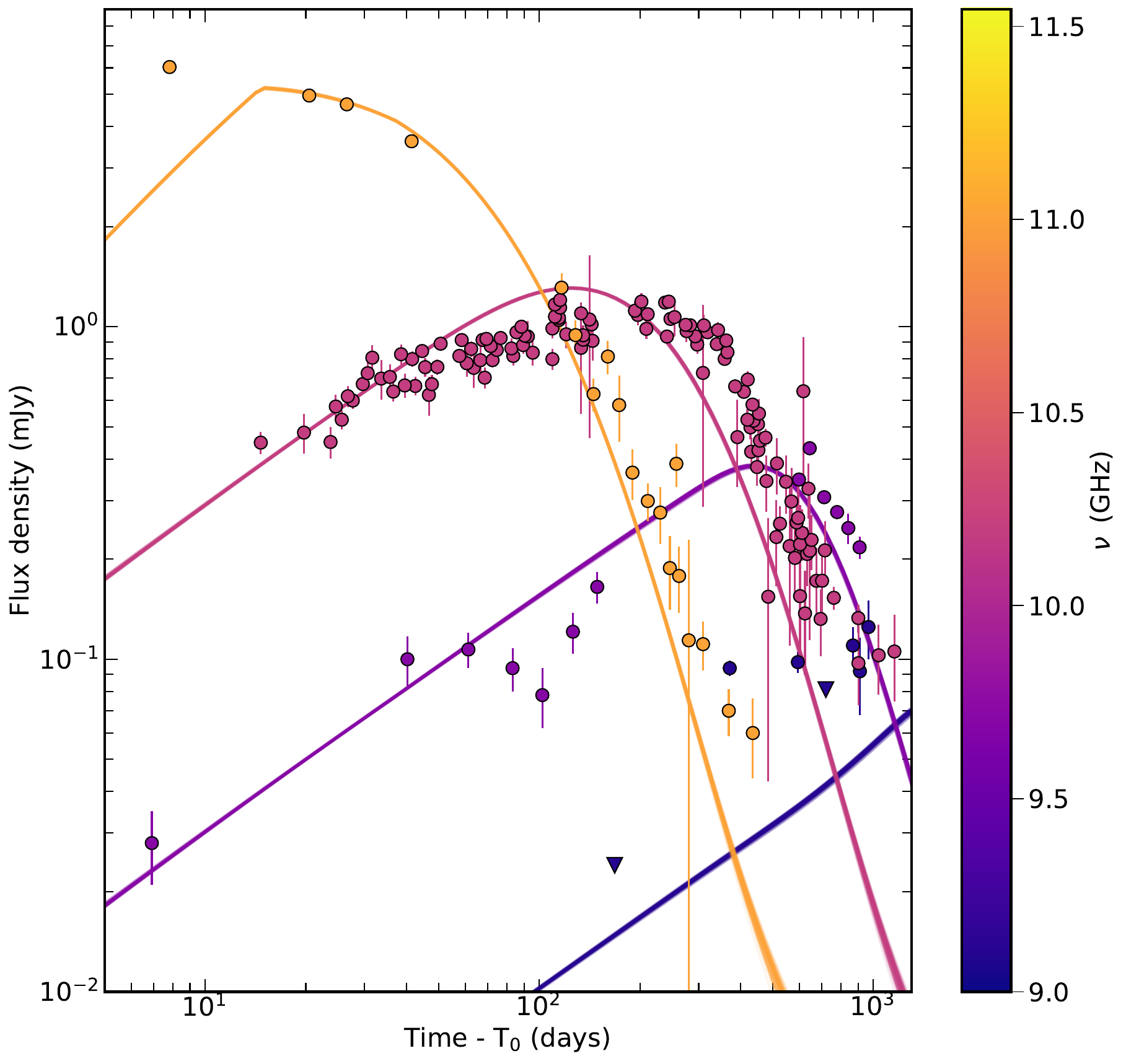}
    \end{subfigure}%
    \begin{subfigure}%[]{}
        \centering
        \includegraphics[width = \columnwidth]{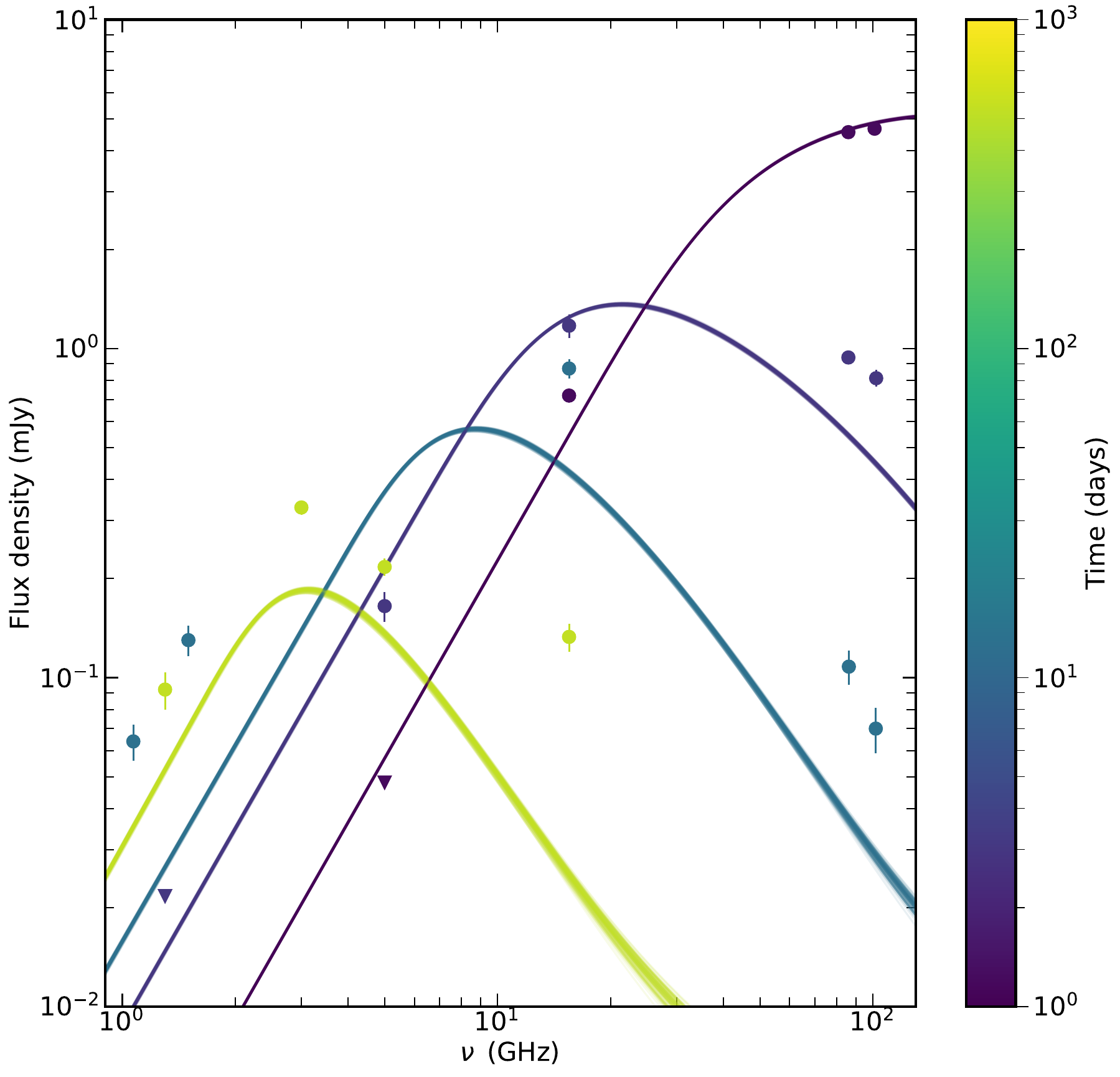}
    \end{subfigure}
    \caption{\textit{Top}: The spherical outflow model fit (Section \ref{subsec:thermal_non_thermal_sph}) the data overlaid on the 1.28, 5, 15.5 and 101.75\,GHz light curves. \textit{Bottom}: The spherical outflow model fit to the data overlaid on the SEDs at $\sim$24, 160, 370 and 910\,days post-discovery. The model tends to underpredict the earliest high frequency and late time 1.28\,GHz data, however this is remains our best fit model. }
    \label{fig:spherical}
\end{figure}

We fit the model described to the entire data set \citep{2022Natur.612..430A, 2023MNRAS.521..389R} using \textsc{emcee} \citep{emcee}. Motivated by theory and particle-in-cell simulations, we fix the microphysical parameters and fit for the hydrodynamics of the outflow. We fix $\epsilon_{\rm{T}} = 0.4$ \citep{2013ApJ...771...54S, 2024PhRvL.132z5201V} and $\epsilon_{\rm{e}} = 0.1$ in the case of a relativistic shock \citep{2013ApJ...771...54S, 2017MNRAS.472.3161B}. We also fix the fraction of the energy in the magnetic fields $\epsilon_{\rm{B}} = 0.1$ \citep{2013ApJ...771...54S} which is similar to $\epsilon_{\rm{e}} = 0.1$ because the magnetic fields are self-amplified by the shock, and the instabilities that amplify the magnetic field would saturate once they start feeding back on the shock. This implies that the amplified magnetic field should have approximately comparable (or slightly less) energy than other components of the shock. For the parameters we fit, all had flat priors: 
\begin{itemize}
    \item post shock velocity $(\Gamma\beta)_{\rm{sh}}$ [$0.0$, $10.0$] allowing for some power law deceleration: $\alpha_{\rm{(\Gamma\beta)_{sh}}}$ [$0.0$, $4.0$]), as well as  $l_{\rm{dec}}$ [$0.0$, $8.0$] which relates the instantaneous shock velocity to the shock radius\footnote{The lab frame radius $R = l_{\rm{dec}} \sqrt{1 + (\Gamma\beta)_{\rm{sh}}^{2}} (\Gamma\beta)_{\rm{sh}} c (1+z) t$ where $t$ is the lab frame time.},
    \item the number density of the environment $\log(n)$ [$-3.0$, $0.0$]): assuming a power law profile away from the black hole $k$ [$0.0$, $4.0$],
    \item the non-thermal electron energy spectral index $p$ [$2.0$, $3.5$].
\end{itemize} 
Our \textsc{emcee} run used 32 walkers and ran the chains for ${>}$10000 steps to increase the chance of convergence. Figure \ref{fig:spherical} shows the model with the posteriors of our \textsc{emcee} run overlaid our data along side that from \citet{2022Natur.612..430A} and \citet{2023MNRAS.521..389R}. %Also shown is corresponding the corner plot in Figure \ref{fig:corner_thermal_non_thermal_sph}. 
The top panel of Figure \ref{fig:spherical} shows the light curves at 101.75, 15.5, 5 and 1.28\,GHz. The model agrees well with the shape of the 101.75, 15.5 and 5\,GHz light curves. However, it struggles to match with the earliest high frequency data and cannot recreate the lowest (1.28\,GHz) light curve. The bottom panel of Figure \ref{fig:spherical} shows the SEDs at 24, 160, 370 and 910 days with the model evaluated at the same time steps. The model reproduces the observed steep spectral index at sub-mm frequencies and shows evidence for the spectrum becoming more shallow at later times. It is demonstrated more clearly here that  the model tends to underpredict the flux density both at high frequencies and late times. 

We find that the shock velocity follows:

$$(\Gamma\beta)_{\rm{sh}} = 1.795 ^{+0.002}_{-0.003} (t/45\rm{d}) ^{-0.288\pm0.001}$$

\noindent the number density of the circumnuclear environment has a profile with 

$$n = 191\pm2 (R/R_{45}) ^{-1.795^{+0.002}_{-0.003}} \rm{cm}^{-3}$$

\noindent where 
%$R_{45}$ 
$R_{45} \approx 9.4 \times 10^{17}\,{\rm cm}$ 
is the shock radius at 45\,days and we obtain a non-thermal electron spectral index of 
$$p = 2.79\pm0.06,$$

and 

$$l_{\mathrm{dec}} = 1.001_{-0.001}^{+0.003}.$$

\noindent where the model favours the lowest possible values of $l_{\mathrm{dec}}$.

\subsection{Jetted outflow with thermal and non-thermal emitting particles}\label{subsec:thermal_non_thermal_jet}

The model presented in \citet{2024ApJ...977..134M} considers a spherical outflow or a jetted system for which the edge of the jet is not be observed. Here we build on \citet{2024ApJ...977..134M} and consider a jet, a blastwave that starts off highly relativistic following a Blandford--McKee solution, a jet break and then a transition into a Sedov--Taylor regime. We also consider the effect of observing the edge of the jet on the radio and submillimeter counterpart \citep{1999ApJ...519L..17S, 2013NewAR..57..141G}.

As with the spherical model, we used \textsc{emcee} to test the model against the data \citep{emcee}. The jet model is parameterized differently, in addition to $p$ and $n \propto r^{-k}$, we fit for the jet evolution  which is dictated by the kinetic energy ($E_{\rm kin}$) adjusted for the jet's opening angle $\theta_{\rm{j}}$, as well as $\epsilon_{e}$, $\epsilon_{B}$, $\epsilon_{T}$. We fit for the microphysical parameters because in the initial testing of fiducial values were not able to reproduce the rise of the light curves. We used flat priors on all parameters: 

\begin{itemize}
    \item for the microphysical parameters $\log(\epsilon_B)$ [$-6.0$, $0.0$],  $\log(\epsilon_e)$ [$-6.0$, $0.0$] and $\log(\epsilon_T)$ [$-6.0$, $0.0$],
    \item the isotropic equivalent kinetic energy $\log(E_{\rm{kin}})$ [$50$, $55$] (erg)
    \item the jet opening angle $\theta_{j}$ [$0.0$, $30$] (deg)
    \item the number density of the environment $\log(n)$ [$-4.0$, $10.0$] and its profile, assuming a power law profile away from the black hole $k$ [$0.0$, $4.0$],
    \item and the non-thermal electron energy spectral index p [$2.0$, $3.5$].
\end{itemize}

\begin{figure}
    \centering
    \begin{subfigure}[]{}
        \centering
        \includegraphics[width = \columnwidth]{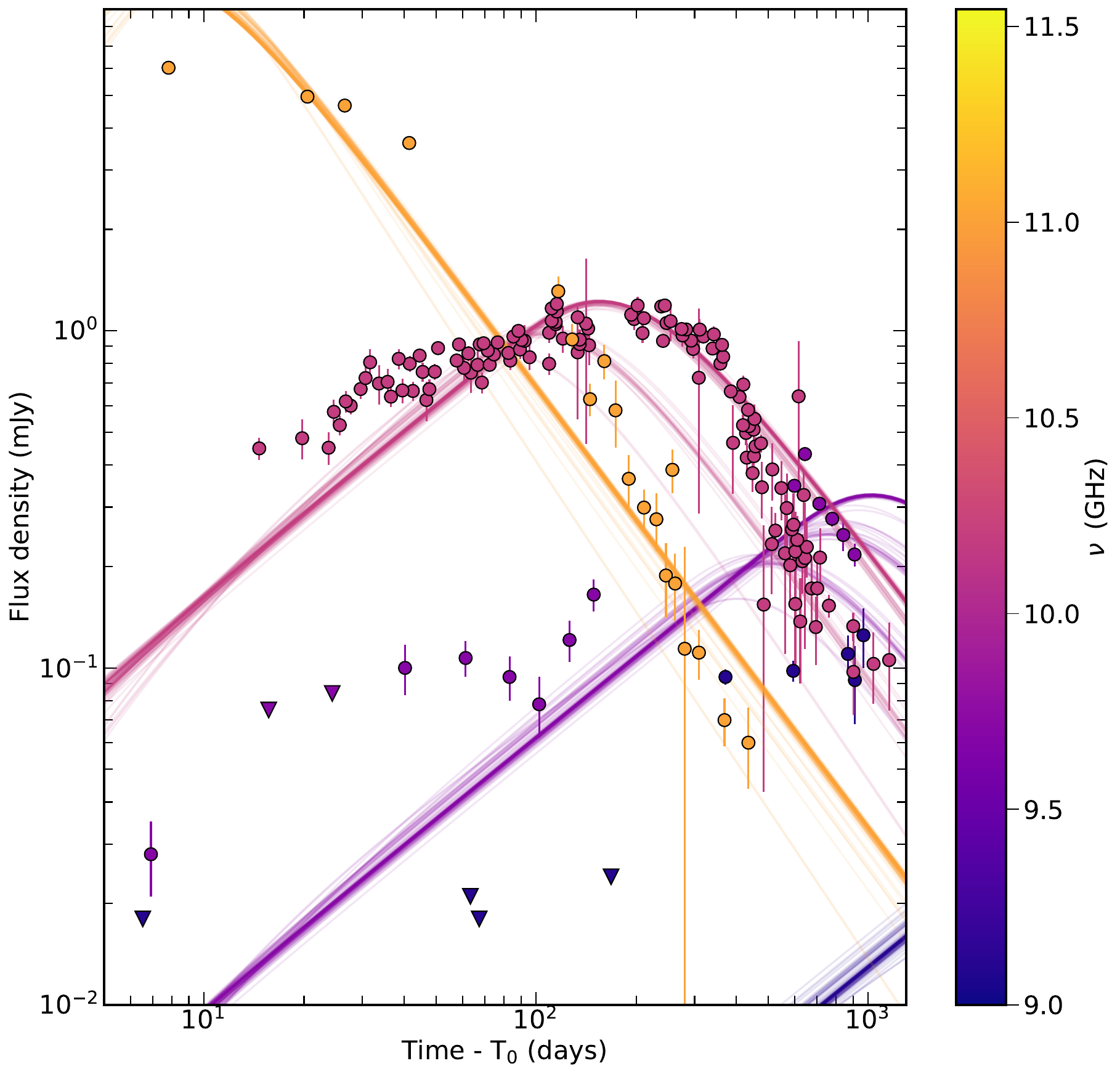}
    \end{subfigure}%
    \begin{subfigure}[]{}
        \centering
        \includegraphics[width = \columnwidth]{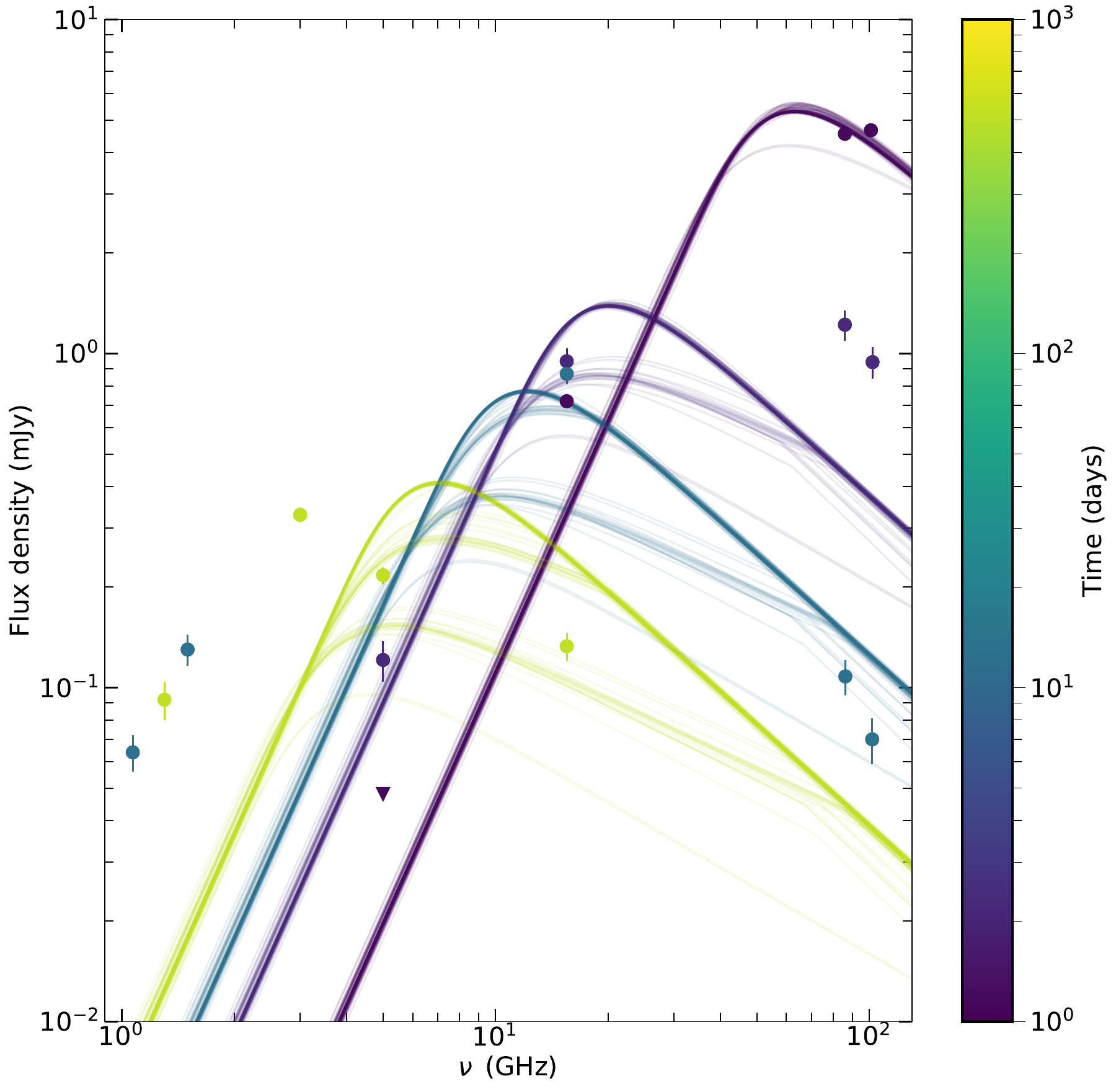}
    \end{subfigure}
    \caption{\textit{Top}: The jetted outflow model (Section \ref{subsec:thermal_non_thermal_jet}) considering both non-thermal and thermal electrons to the data overlaid on the 1.28, 5, 15.5 and 101.75\,GHz light curves. \textit{Bottom}: The jetted outflow model fit to the data overlaid on the SEDs. Given the scatter in the model, we only model the SEDs at $\sim$24, 160, 370 and 910\,days post-discovery to prevent over cluttering the Figure. The jetted model cannot explain the steep spectral index, the high frequency turn over in the first $\sim10$\,days post-discovery or the low frequency emission.}
    \label{fig:jet}
\end{figure}

The results of our \textsc{emcee} run are presented in Figure \ref{fig:jet} which shows light curves and SEDs produced by randomly sampling the posterior (post burn-in) 100 times. The walkers in the \textsc{emcee} run struggled to converge even when they ran for $\sim10^{7}$ steps. The main issue of this jet model is that the spectral evolution is not fast enough to catch the peak at each frequency. This is highlighted particularly in the panel of Figure \ref{fig:jet}, with the peak of the 101.75\,GHz light curve which is not reproduced and the 1.28\,GHz data, where the predicted flux density is a factor of 10 off from the observed emission. We also find no change in spectral index within the model, at all times the high frequency spectral index is too shallow to match the data (see high frequency side of the lower panel of Figure \ref{fig:jet}). Overall, comparison of the model and our data shows that an on-axis relativistic jet cannot reproduce the observed behavior.  

From the jet model, we find that kinetic energy in the jet is $$E_{K}=10^{52.5_{-0.2}^{+0.1}}\rm{erg},$$ and the density and the corresponding density profile follow $$n = 10^{7.1_{-0.4}^{+0.3}} (R/10^{17})^{-0.02_{-0.02}^{+0.01}}\rm{cm}^{-3}.$$ The jet requires an opening angle of $$\theta_{j}=7_{-3}^{+2~\circ}$$ and the electron energy spectral index is $$p = 2.02_{-0.01}^{+0.02}$$ \noindent which is as low as our set priors would allow. Unlike in the spherical scenario, the summation of $\epsilon_{e}$, $\epsilon_{T}$ and $\epsilon_{B}$ is less than one ($0.8\pm0.1$): $$\epsilon_{e} = 0.8\pm0.1, \epsilon_{T} = (3_{-1}^{+3})\times10^{-4}, \epsilon_{B} = (2\pm1)\times10^{-4}.$$ 

We compare the physical parameters derived from the model to other radio-detected TDEs and find that the density and profile is very different to other results in the literature \citep[e.g.][]{2016ApJ...819L..25A,2017ApJ...837..153A,2018ApJ...854...86E,2021ApJ...919..127C,2024MNRAS.528.7123G}, where the density profiles of the respective circumnuclear environments fall off between $r^{-1}$ and $r^{-2}$ as opposed to the flat density profile with a much higher normalization.

Given the poor fit to the data and the very high densities inferred, we conclude that the radio and sub-mm emission from AT2022cmc cannot be described by the jet model described here.

\section{Discussion} \label{sec:discussion}

Our observations of AT2022cmc up to 1000\,days post-discovery reveal a long-lasting radio and sub-mm counterpart. The light curves in each band show a variety of behavior with steeper decays at higher frequencies ($>$15\,GHz) than lower frequencies (1.28-5\,GHz). The broadband SEDs show an optically thick ($\nu^{\sim2}$) branch that transitions to a steep optically-thin branch ($\nu^{-1.4}$). After $\sim$600\,days the steep optically thin spectrum becomes shallower ($\nu^{-0.7}$). Based on the steepness of the sub-mm spectral index, we infer the presence of a thermal electron population, in addition to a non-thermal population which begins to dominate at later times.
Such a change in the spectral index over time is a hallmark expectation in models that include thermal electrons \citep[e.g.,][]{2021ApJ...923L..14M}.

\subsection{Outflow velocity and density profile}

The shape of the SED is a strong function of the shock proper velocity ($\Gamma\beta$)$_{\rm{sh}}$ \citep{2024ApJ...977..134M}. At the highest values of ($\Gamma\beta$)$_{\rm{sh}}$ ($\gtrsim 2$), the peak of the SED is optically thin. This scenario is observed in gamma-ray burst afterglows which have bulk Lorentz factors of 10s to 100s \citep{2018A&A...609A.112G}. \citet{2022Natur.612..430A} fit a series of synchrotron spectral to their radio and sub-mm SEDs spanning 5--45\,days. At 11 and 20\,days they find that the peaks of the SEDs are optically thin, which according to \citet{2024ApJ...977..134M} would require ($\Gamma\beta$)$_{\rm{sh}} > 2$, leading to possible agreement \citet{2023MNRAS.521..389R}. At slightly lower velocities ($0.3 \lesssim (\Gamma\beta$)$_{\rm{sh}} \lesssim 2$), the SED peak becomes optically thick and the emission is expected to be dominated by thermal electrons (if they are present) which creates a steep spectral slope just above the peak. Such a steep slope was observed in the sub-mm counterpart to AT2022cmc until $370$\,days when sub-mm observations ceased, indicating both a significant thermal electron population and that the outflow is mildly relativistic but slower than at $\sim10$\,days \citet{2022Natur.612..430A}. We find there is an agreement between the expected SED slope and the observed emission. At ever lower velocities (($\Gamma\beta$)$_{\rm{sh}} \lesssim 0.3$), the peak of the SED is still optically thick but dominated by non-thermal electrons \citep{1998ApJ...499..810C}. Above the peak frequency, the spectral index shallows, as we observed after $\sim500$\,days. Qualitative comparison of our SEDs to those derived in different velocity regimes in \citet{2024ApJ...977..134M} show that there must be deceleration from an initially-relativistic outflow. The fitting our model to the data finds that the outflow is decelerating ($\Gamma\beta$)$_{\rm{sh}} = 1.795 (\rm{t}/45\rm{d})^{-0.288}$, and shows the clear change in spectral index despite the later-time SEDs not being fit well by the spherical model.

Figure \ref{fig:Energy_beta} shows the post-shock energy ($U$) against the shock velocity ($(\Gamma\beta)_{\rm{sh}}$)  for AT2022cmc (from our spherical model, light orange solid circles with navy edges), thermal TDEs (gold squares) and \textit{Swift} J1644 (dark orange stars) as well as FBOTs (fast blue optical transients) and three difference classes of supernovae (dark circles). The thermal TDEs sit in a lower velocity, lower energy region of the parameter space alongside the supernovae, away from \textit{Swift} J1644 and AT2022cmc. In terms of $(\Gamma\beta)_{\rm{sh}}$, \textit{Swift} J1644 and AT2022cmc fall in the same region of the parameter space. Comparison of our results to both \citet{2024ApJ...965...39Y} and \citet{2023MNRAS.522.4028M} finds that we obtain marginally large proper velocity values over the same time frame. Conversely, our energy measurements for AT2022cmc ($\sim10^{52}$\,erg) are nearly 2 orders of magnitude higher than found in \citet[][]{2024ApJ...965...39Y} who found $(3-5)\times10^{50}$\,erg and a factor of a few lower than the isotropic equivalent energy calculated by \citet[][$\sim9\times10^{52}$\,erg]{2023MNRAS.522.4028M}. %however, we obtain very similar outflow velocity ranges of $\sim2-3$ between $\sim5-40$\,days post-discovery. The discrency due to not including the contribution from the thermal electrons. % Despite the disagreement with \citet{2023MNRAS.521..389R, 2023NatAs...7...88P} we do agree with the findings of \citet{2024ApJ...965...39Y}. 
In the case of \citet{2024ApJ...965...39Y}, we hypothesise that the difference in the total energy inferred between our study and \citet{2024ApJ...965...39Y} can be attributed to the inclusion of thermal electrons. In our spherical model, we assume $\epsilon_{\rm{e}} =0.1$ and $\epsilon_{\rm{T}} = 0.4$, i.e., the thermal electron population has four times as much energy as the thermal electrons. Therefore, if only non-thermal electrons are identified and modelled the total energy inferred will be sustantially lower. 

We have demonstrated both quantitatively and qualitatively from the SED that the outflow associated with AT2022cmc is decelerating. The deceleration of the jet is dictated by the density and density profile of the environment the outflow is propagating through. A steeper density profile corresponds to a slower deceleration. Figure \ref{fig:N_R} shows the number density of the circumnuclear environment as a function of distance from the supermassive black hole for a sample of TDEs from \citet{2016ApJ...819L..25A,2017ApJ...837..153A,2018ApJ...854...86E,2021ApJ...919..127C,2022MNRAS.511.5328G,2023MNRAS.518..847G,2023MNRAS.522.5084G,2024MNRAS.528.7123G}. Overlaid is the density profile of AT2022cmc derived from the spherical model (Section \ref{subsec:thermal_non_thermal_sph}) evaluated at 24, 160, 370 and 910\,days post-discovery. We find that the density and profile of the environment that AT2022cmc's outflow is propagating through is consistent with the rest of the TDE population. In addition, we find a similar density and profile to \citet{2023MNRAS.522.4028M}'s analysis of \citet{2022Natur.612..430A}'s of AT2022cmc data, using a jet model with an opening angle of 0.1\,radians. The density profile for the spherical outflow model is completely consistent ($R^{-1.795^{+0.002}_{-0.003}}$) with that seen in other TDEs ($R^{-1.5}$-- $R^{-2}$).

It is interesting to note that the best-fit parameters inferred for the outflow deceleration $\alpha_{\Gamma\beta_{\rm sh}} \approx 0.3$ and the external density profile $k \approx 1.8$ are in rough agreement with theoretical closure relations. A spherically-symmetric energy-conserving (adiabatic) blast-wave expanding into a power-law density profile, $n \propto r^{-k}$,
produces a shock proper-velocity that depends on shock radius as $(\Gamma\beta)_{\rm sh} \propto r^{-(k-3)/2}$.
This scaling is correct in both the non-relativistic and ultra-relativistic regimes where it correctly describes the Sedov-Taylor and Blandford-McKee solutions, respectively. These solutions are also correct for jetted or conical outflows so long as the jet does not expand laterally (i.e. accurate before the jet-break time in relativistic outflows). Our model is set up assuming that the proper-velocity of the shock has a power-law temporal dependence rather than a power-law dependence with radius, $$(\Gamma\beta)_{\rm sh} \propto t^{-\alpha_{\Gamma\beta_{\rm sh}}}$$ but we note that this can be related to the shock radius, giving
$$\frac{\rm{d}\ln (\Gamma\beta)_{\rm{sh}}}{\rm{d}\ln{r}} = \left[ 1 - \alpha_{(\Gamma\beta)_{\rm sh}}^{-1} + (\Gamma\beta)_{\rm sh} / \sqrt{1 + (\Gamma\beta)_{\rm sh}^2} \right]^{-1}.$$
For an energy-conserving spherical (or conical without lateral expansion) solution as described above with $k = 1.8$ (as inferred), and for $(\Gamma\beta)_{\rm sh} \sim 1$ (relevant for most times of interest), 
this closure relation implies that
$\alpha_{(\Gamma\beta)_{\rm sh}} \approx 0.3$, which is consistent with the inferred $\alpha_{\Gamma\beta_{\rm sh}}$ value obtained by our \textsc{emcee} fit. There is no need that this be the case, as we did not enforce any correlation between $k$ and $\alpha_{(\Gamma\beta)_{\rm sh}}$ in our fitting procedure, so the two parameters are formally independent in our model.
The fact that the inferred parameters satisfy this closure relation therefore adds confidence in the physical plausibility of our fit. It also implies that the outflow is well within the energy-conserving phase of its hydrodynamic evolution, as opposed to being in the `ejecta dominated' phase where the original outflow distribution still has a significant impact \citep[e.g.][]{1999ApJS..120..299T}.

\subsection{Outflow Geometry}

\begin{figure}[h]
    \centering
    \includegraphics[width=\columnwidth]{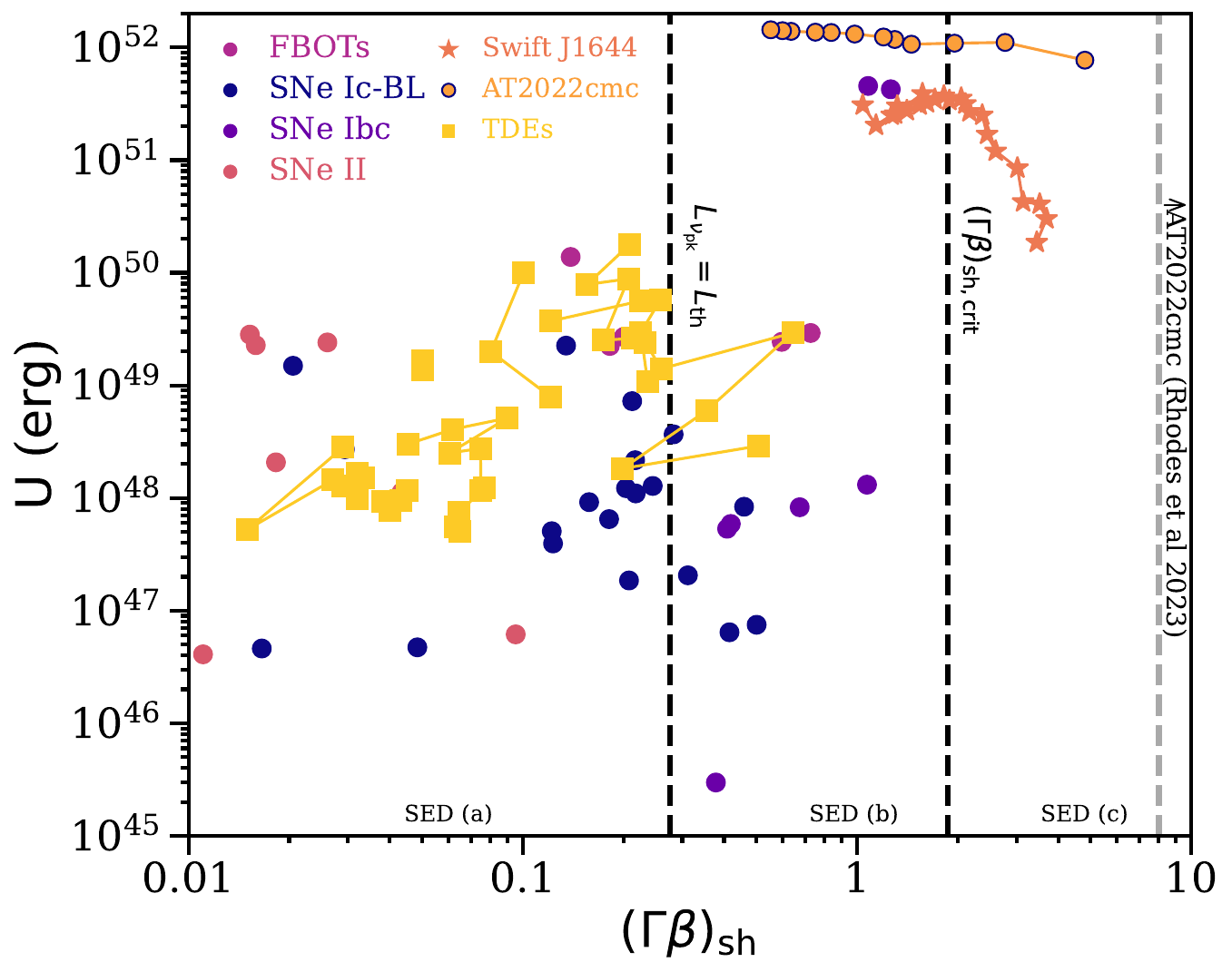}
    \caption{Post-shock energy versus shock velocity inferred from peak properties of synchrotron-powered transients including \textit{Swift} J1644 and AT2022cmc \citep{2018ApJ...854...86E,2024ApJ...977..134M}. We find that AT2022cmc has similar outflow velocity compared to \textit{Swift} J1644 and a higher total energy content. The maximum outflow velocity we derive is not high enough to match the lower limit found in \citet{2023MNRAS.521..389R}.}
    \label{fig:Energy_beta}
\end{figure}

\begin{figure}[h]
    \centering
    \includegraphics[width=\columnwidth]{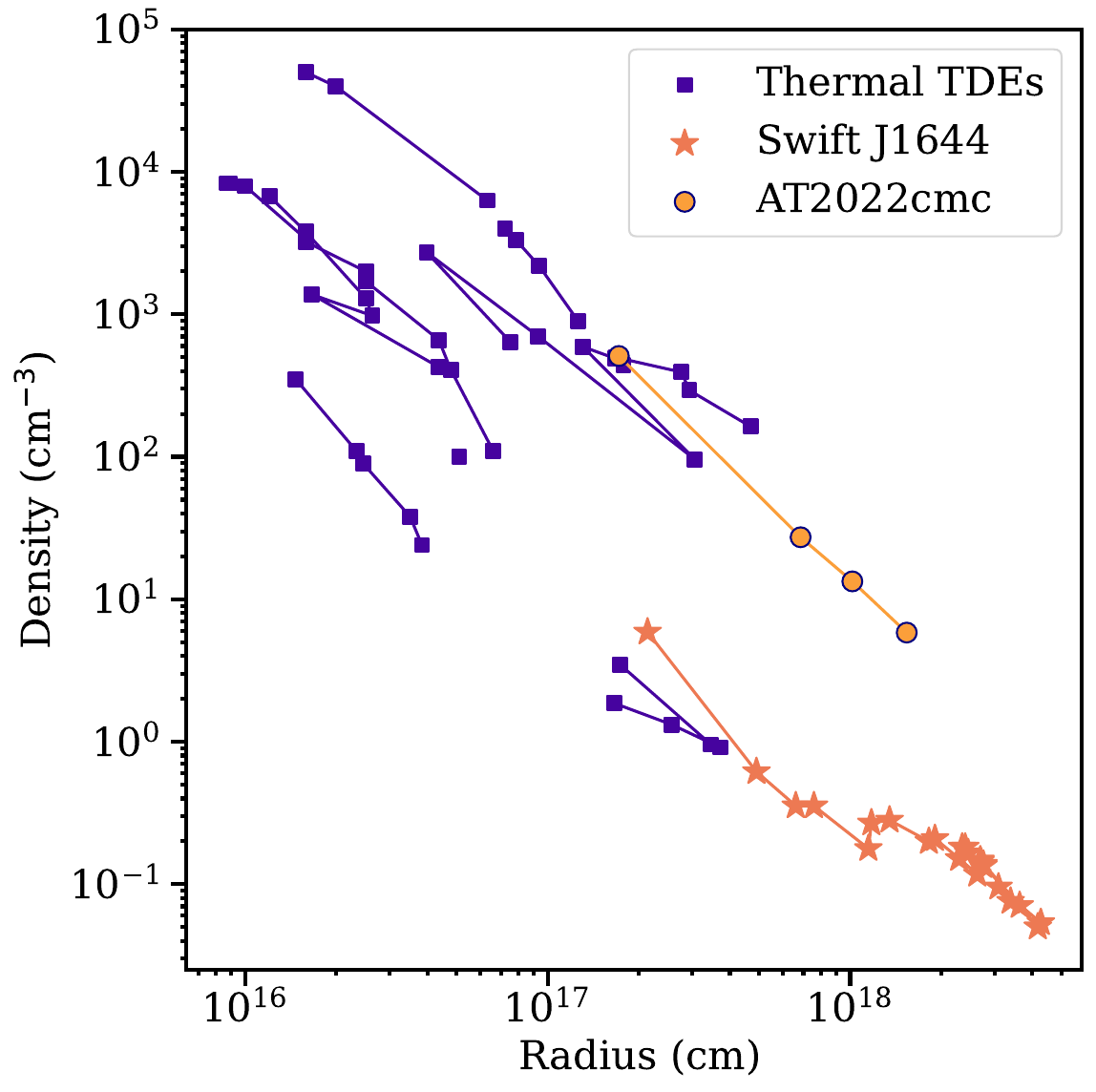}
        \caption{ The number density of the circumnuclear environment as a function of distance from the supermassive black hole for a sample of radio-detected TDEs including the other well-studied relativistic TDE \textit{Swift} J1644 \citep{2016ApJ...819L..25A,2017ApJ...837..153A,2018ApJ...854...86E,2021ApJ...919..127C,2022MNRAS.511.5328G,2023MNRAS.518..847G,2023MNRAS.522.5084G,2024MNRAS.528.7123G}. Also shown is the density profile for AT2022cmc which shows the same density profile as the rest of the TDE population. }
    \label{fig:N_R}
\end{figure}

The jet models presented in Sections \ref{subsec:non_thermal_jet} and \ref{subsec:thermal_non_thermal_jet}, are equivalent to the spherical model (Section \ref{subsec:thermal_non_thermal_sph}) until the beaming angle (1/$\Gamma$) is greater than the jet opening angle. One conclusion that can be drawn from how well the spherical model fits the observations is that, \textit{if} a single-component top hat jet (one with no lateral structure) is producing all the radio and sub-mm emission then the edges of a jet have not been observed yet. The jet would still be moving fast enough that the beaming angle of the jet is larger than 1/$\Gamma$. However, as shown in Figure \ref{fig:Energy_beta}, by 900\,days post-discovery $\Gamma\beta\sim0.8$, which corresponds to a beaming angle of $\sim45^{\circ}$. So if a narrow jet was present a jet break should have been visible. 

A jet break may have been hidden in a case where there is radial structure to the outflow such that the jet does not have a traditional `top hat' shape. \citet{2024ApJ...974..162Y} suggested that the multi-wavelength emission from AT2022cmc could be explained by a combination of a narrow, fast jet and a wider, slow jet, where each component produces its own emission signature. Lateral structure in the jet could explain not only the lack of jet break features but also the deviations between the model and data at later times. The top panel of Figure \ref{fig:spherical} shows that the model is under-predicting the 1.28\,GHz light curve by a factor of two. It maybe possible that a second component is starting to contribute to the radio SED at later times \citep{2024ApJ...963...66Z}. Continued monitoring of AT2022cmc at cm-wavelengths will be needed to confirm or rule out this possibility.

\subsection{Comparison to other radio transients}

Radio observations of TDEs have allowed for a large variety of outflow properties to be inferred \citep{2020SSRv..216...81A,2024ApJ...971..185C, 2025arXiv250408426G}. There is a wide range in luminosity and variability-timescale parameter space within which relativistic TDEs are the most radio-luminous. Within the growing sample of radio-detected TDEs, there are now a number of relativistic TDEs in the literature both on-axis and off-axis \citep[e.g., ][]{2011Natur.476..425Z,2022Natur.612..430A,2024MNRAS.527.7672S}. However, only \textit{Swift} J1644 has sufficient radio coverage on appropriate timescales for us to make detailed comparisons with AT2022cmc.

\begin{figure}%[h]
    \centering
    \vspace{0.5cm}
    \includegraphics[width=\columnwidth]{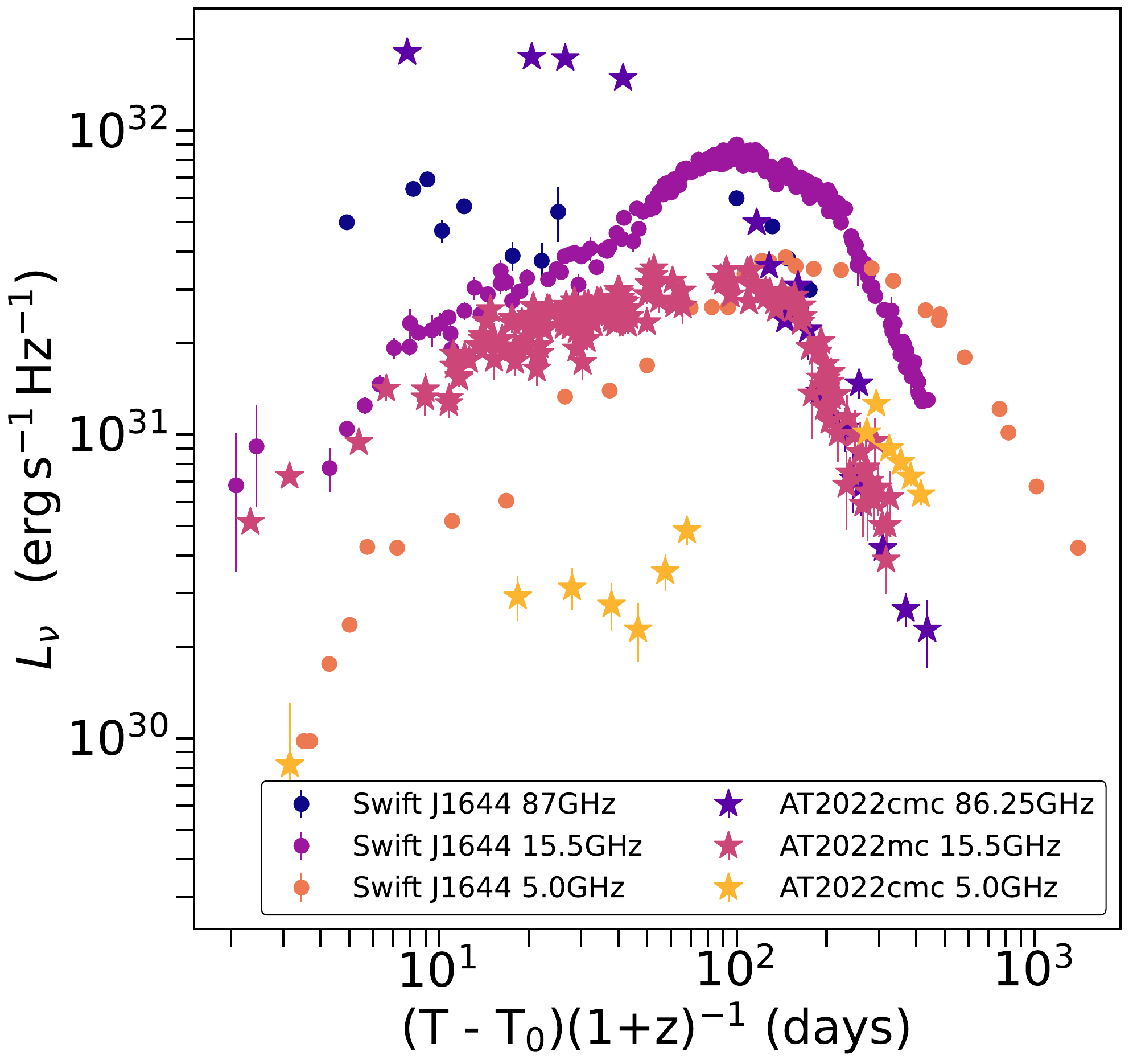}
    \caption{The 86/87, 15 and 5\,GHz luminosity light curves for both AT2022cmc and \textit{Swift} J1644+57. While \textit{Swift} J1644 is systematically more luminous than AT2022cmc, the two events show remarkably similar light curves but \textit{Swift} J1644 has has yet to be intepreted within a thermal$+$non-thermal scenario.}
    \label{fig:Lum}
\end{figure}

In addition to the physical parameter comparisons that have already been made, we directly compare the observational data. Figure \ref{fig:Lum} shows the 86--87, 15.5 and 5\,GHz luminosity light curves for both \textit{Swift} J1644 (circles) and AT2022cmc (stars) \citep{2013ApJ...767..152Z, 2012ApJ...748...36B}. \textit{Swift} J1644 is systematically more luminous than AT2022cmc for most of the observing period, but the general shape of the light curves for each event are the same at a particular frequency. We highlight two areas where there are distinct differences between the two events. First is the high frequency light curves of AT2022cmc and \textit{Swift} J1644 differ by a nearly a factor of 3 until the turn over at around 100\,days. After this they follow the same decay at the same luminosity, whereas at other frequencies there is a clear offset in their luminosities. Second, we observe a rising component in the 15.5\,GHz light curves. The rise in the \textit{Swift} J1644 light curve has more structure whereas the AT2022cmc data is well described by a single power law. We highlight these two areas in particular because we find, at least initially, they best explain the differences in the subsequent interpretations of the data. 

\begin{figure}
    \centering
    \includegraphics[width=\columnwidth]{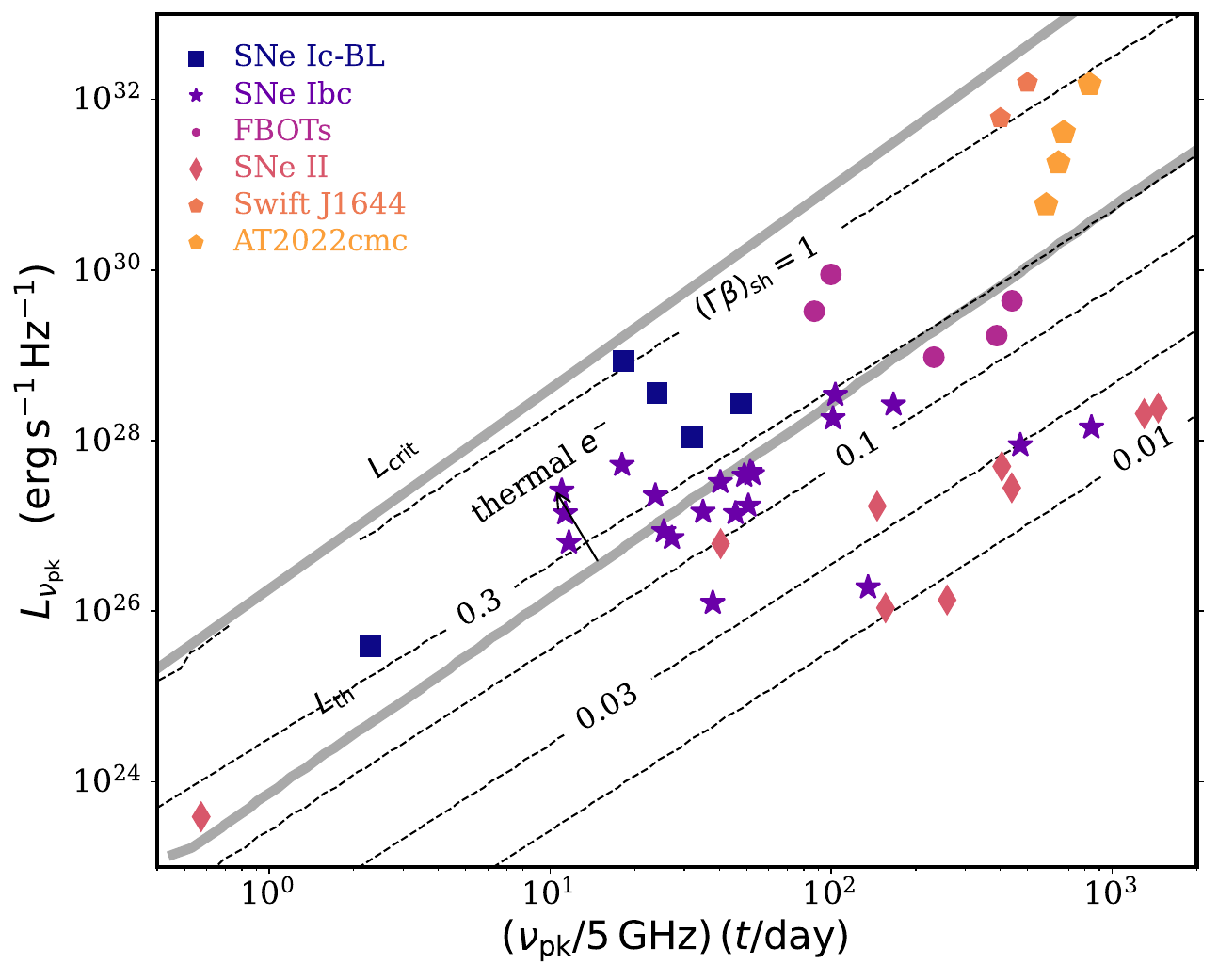}
    \caption{Radio luminosity -- peak frequency parameter space plot \citep{1998ApJ...499..810C} recreated from figure 2 of \citet{2024ApJ...977..134M}. The black dashed lines denote lines of constant $(\Gamma\beta)_{\rm sh}$. The bottom grey line indicates the region above which emission at the peak frequency is dominated by thermal electrons. The top grey line labelled $L_{\rm crit}$ shows a theoretical upper bound for points within this parameter space \citep{2024ApJ...977..134M}. We have plotted points corresponding to AT2022cmc at 24, 260, 370 and 910\,days.}
    \label{fig:fig2_margalit}
\end{figure}

In \textit{Swift} J1644, there has been no search for an thermal electron population, instead, there have been a range of explanations including gamma-ray burst-like afterglows, inverse Compton cooling, and synchrotron$+$synchrotron self-Compton from different emitting regions \citep{2012MNRAS.420.3528M,  2013MNRAS.434.3078K, 2015ApJ...798...13L, 2016MNRAS.460..396C}. A thermal$+$non-thermal electron model has not yet been applied to the radio and sub-mm counterpart to \textit{Swift} J1644. However, we note that in \citet{2021ApJ...908..125C}, SED modelling of the radio counterpart to \textit{Swift} J1644 found that the high frequency spectral index was quite steep favouring $\gamma \approx -1$ ($p = 3$) when considering a non-thermal electron spectrum only. Such a steep spectrum is not unheard of in non-thermal scenarios \citep[e.g.][]{2012ApJ...752...17W} however a thermal$+$non-thermal scenario could also be possible. A thermal$+$non-thermal scenario is reinforced when examining the positions of both AT2022cmc and \textit{Swift} J1644 in Figure \ref{fig:fig2_margalit}. Figure \ref{fig:fig2_margalit} shows the peak radio luminosity plotted against the product of the frequency at which the SED peaks and time at which the SED was measured \citep[e.g.,][]{1998ApJ...499..810C,2024ApJ...977..134M}. The region between the two grey diagonal lines indicates the part of the parameter space where thermal electrons are expected to dominate at the peak frequency ($\nu_{\rm{pk}}$). Ignoring the contribution of the thermal electrons, in this region, would mean that one derives systematically higher densities and shock velocities than when they are included. Both relativistic TDEs sit between the two thick grey lines. Therefore, one might expect other relativistic TDE systems to exhibit signatures of thermal electrons. We encourage a remodelling of the \textit{Swift} J1644 dataset to include thermal electrons.

\subsection{Implications for future observing campaigns}

We have demonstrated using AT2022cmc that sub-mm observations are vital for a better understanding of the jet physics of TDEs. The sub-mm observations give us the longest temporal baseline over which to measure the optically thin spectral index. Such a measurement is the only way to determine the relative fraction of thermal and non-thermal electrons. By combining early time ($<$100\,days) sub-mm data with late-time cm-wavelength data we can measure the change in the optically thin spectral index as the jet evolves, and track the transition from a thermal to non-thermal spectrum. From AT2022cmc, we see that at early times the thermal electrons dominate, but by the time the spectral peak reaches cm-wavelengths, the non-thermal population dominates. Therefore, in future TDE studies, to perform the most comprehensive physically complete modelling, sub-mm observations will be vital.

The discovery of a second relativistic TDE in real time has enabled a detailed study of the broadband emission that would otherwise not be possible and allows us to fill in more of the radio transient parameter space at the highest radio luminosities \citep[e.g. figure 4 of ][]{2024ApJ...969L..14G}. As the Vera C. Rubin observatory comes online, the detection rate of TDEs (and hopefully relativistic TDEs) will increase by several orders of magnitude \citep{2020ApJ...890...73B} providing more opportunities for sub-mm and cm-wavelength follow up and monitoring which we can use to understand the diversity in relativistic TDE parameter space. 

With the launch of Einstein Probe and SVOM, the high-energy astrophysics community are now exploring a new population of jetted sources. Many of the newly discovered Fast X-ray Transients (FXTs, flashes of soft X-ray emission) come with optical and/or radio counterparts that sit in a similar region of luminosity/variability-timescale parameter space to both GRBs and relativistic TDEs. To fully understand the origins of FXTs and other yet-to-be-discovered high energy transients, we need to better understand these known transients. Only then will it be possible to determine whether these new classes of transients are GRBs or relativistic TDEs or some other type of cataclysmic event.

\section{Conclusions} \label{sec:conc}

We have presented radio and sub-mm observations of the fifth relativistic TDE AT2022cmc between 100 and 1000\,days post-discovery. Our observations show a long lasting counterpart which corresponds to an expanding emitting region transitioning from optically thick to thin. Compared to other jetted transients, the observations presented here differ most in the post-peak spectral index. The high frequency spectral indices we measure are relatively steep, and we test the possibility that this originates from an additional component of synchrotron-emitting thermal electrons. We test three different models against our data:

\begin{enumerate}
\item an entirely non-thermal GRB afterglow-like model; 
\item a spherical outflow model that considers both non-thermal and thermal electrons from \citet{2024ApJ...977..134M};
\item a conical outflow model that considers both non-thermal and thermal electrons.

\end{enumerate}

We find that the spherical outflow model provides the best fit to the data. We compare the energetics, shock velocity evolution and circumnuclear environmental properties to that of other TDEs (Figures \ref{fig:Energy_beta} and \ref{fig:N_R}) and find that the outflow in AT2022cmc follows a similar density profile with velocity in agreement to \textit{Swift} J1644 but is more energetic by a factor of $\sim$2. 

We also compare the light curve and spectral properties of AT2022cmc and \textit{Swift} J1644. Interpretations of \textit{Swift} J1644 have only considered non-thermal synchrotron models, despite the clear spectral and temporal similarities between the two events. Therefore we encourage a reanalysis of \textit{Swift} J1644 within a thermal$+$non-thermal framework. Given their similar positions in a comparison between the radio luminosity and peak frequency (Figure \ref{fig:fig2_margalit}), we expect that fitting a thermal$+$non-thermal electron model to the data should provide a good fit.

By identifying a thermal electron population in a jetted transient, this work has opened up a new dimension within which to study black hole jets. To date, black hole transients containing jets, such as gamma-ray bursts or X-ray binaries, have considered only non-thermal electrons (parametrized through $\epsilon_{e}$ and $\epsilon_{B}$). We have demonstrated the shock physics in these relativistic transients is more complex but also that the observations can be interpreted using more complete models. By including thermal electrons in the modelling of future jetted transients, we will improve the accuracy of the derived physical parameters and our global understanding of these transients.

\newpage

\section*{Acknowledgements}

%\begin{acknowledgments}
LR acknowledges support from the Trottier Space Institute Fellowship and from the Canada Excellence Research Chair in Transient Astrophysics (CERC-2022-00009). LR thanks Adelle Goodwin for contributing to Figures \ref{fig:Energy_beta} and \ref{fig:N_R}. RPF acknowledges support from UKRI, The ERC and The Hintze Family Charitable Foundation. A.H. is grateful for the support by the Israel Science Foundation (ISF grant 1679/23) and by the United States-Israel Binational Science Foundation (BSF grant 2020203). This research was supported in part by grant NSF PHY-2309135 to the Kavli Institute for
Theoretical Physics (KITP). D.H. acknowledges funding from the NSERC Arthur B. McDonald Fellowship and Discovery Grant programs and the Canada Research Chairs (CRC) program. The authors acknowledge support from the Centre de recherche en astrophysique du Québec, un regroupement stratégique du FRQNT.

e-MERLIN is a National Facility operated by the University of Manchester at Jodrell Bank Observatory on behalf of STFC.  
This project has received funding from the European Union’s Horizon 2020 research and innovation programme under grant agreement No 101004719. We thank the staff at the Mullard Radio Astronomy Observatory for carrying out observations with the Arcminute Microkelvin Imager Large Array.

\vspace{5mm}
\facilities{AMI--LA, \textit{e}-MERLIN, MeerKAT, NOEMA}


\software{\textsc{python}}



\appendix

\section{Observations} \label{sec:app_obs}


\begin{center}
\begin{longtable}{ccccc}
\caption{A table of all the new observations reported in this paper. Values in the flux density column prefaced with $<$ are 3$\sigma$ upper limits. The times are the start time of each epoch. The uncertainties on the flux density are calculated by adding the statistical error on the fit and a calibration error in quadrature. For AMI--LA, NOEMA and \textit{e}-MERLIN, we use a calibration error of 5\% and for MeerKAT we use 10\%. The epochs denoted with $^{a}$, $^{b}$ and $^{c}$ are the results of concatenating the data collected in March 2024; July and August 2024, and April 2025, respectively. For the concatenated data sets, we provide the central time as opposed to the observing start time.} \label{tab:data} \\

\hline \multicolumn{1}{|c|}{Date(dd-mm-yyyy)} & \multicolumn{1}{c|}{Time (UTC)} & \multicolumn{1}{c|}{Telescope} & \multicolumn{1}{c|}{Central Frequency (GHz)} & \multicolumn{1}{c|}{Flux Density (mJy)} \\ \hline 
\endfirsthead

\multicolumn{5}{c}%
{{\bfseries \tablename\ \thetable{} -- continued from previous page}} \\
\hline \multicolumn{1}{|c|}{Date(dd-mm-yyyy)} & \multicolumn{1}{c|}{Time (UTC)} & \multicolumn{1}{c|}{Telescope} & \multicolumn{1}{c|}{Central Frequency (GHz)} & \multicolumn{1}{c|}{Flux Density (mJy)} \\ \hline 
\endhead

\hline \multicolumn{5}{|r|}{{Continued on next page}} \\ \hline
\endfoot

\hline \hline
\endlastfoot
24/05/2022 & 12:45:00 & \textit{e}-MERLIN &5 &0.08$\pm$0.02\\
31/05/2022 & 20:57:05 & AMI-LA &15.5 &0.80$\pm$0.07 \\
31/05/2022 & 20:57:18 & AMI-LA &15.5 &0.99$\pm$0.08 \\
02/06/2022 & 19:51:22 & AMI-LA &15.5 &1.17$\pm$0.08 \\
02/06/2022 & 19:51:35 & AMI-LA &15.5 &1.07$\pm$0.07 \\
05/06/2022 & 23:01:47 & AMI-LA &15.5 &1.06$\pm$0.09 \\
05/06/2022 & 23:02:14 & AMI-LA &15.5 &1.05$\pm$0.09 \\
06/06/2022 & 20:11:19 & AMI-LA &15.5 &1.14$\pm$0.08 \\
06/06/2022 & 20:11:46 & AMI-LA &15.5 &1.20$\pm$0.07 \\
07/06/2022 & 23:55:00 & NOEMA &86.25 &1.6$\pm$0.2\\
07/06/2022 & 23:55:00 & NOEMA &101.75 &1.3$\pm$0.1\\
11/06/2022 & 19:17:13 & AMI-LA &15.5 &0.95$\pm$0.1 \\
17/06/2022 & 12:00:00 & \textit{e}-MERLIN & 5 &0.12$\pm$0.02\\
19/06/2022 & 18:53:00 & NOEMA &86.25 &1.2$\pm$0.1\\
19/06/2022 & 18:53:00 & NOEMA &101.75 &0.9$\pm$0.1\\
24/06/2022 & 19:02:31 & AMI-LA &15.5 &1.1$\pm$0.1 \\
24/06/2022 & 19:02:59 & AMI-LA &15.5 &$<$1.0 \\
26/06/2022 & 16:56:06 & AMI-LA &15.5 &0.91$\pm$0.06 \\
26/06/2022 & 16:56:27 & AMI-LA &15.5 &0.94$\pm$0.08 \\
02/07/2022 & 18:02:36 & AMI-LA &15.5 &$<$1.8 \\
04/07/2022 & 19:57:56 & AMI-LA &15.5 &1.02$\pm$0.09 \\
05/07/2022 & 18:10:46 & AMI-LA &15.5 &0.91$\pm$0.12 \\
06/07/2022 & 17:25:00 & NOEMA &86.25 &0.78$\pm$0.08\\
06/07/2022 & 17:25:00 & NOEMA &101.75 &0.63$\pm$0.07\\
10/07/2022 & 11:30:00 & \textit{e}-MERLIN &5 &0.17$\pm$0.02\\
21/07/2022 & 19:56:00 & NOEMA &86.25 &0.9$\pm$0.1\\
21/07/2022 & 19:56:00 & NOEMA &101.75 &0.81$\pm$0.09\\
29/07/2022 & 12:59:55 & MeerKAT &1.28 &$<$0.03\\
03/08/2022 & 19:21:00 & NOEMA &86.25 &0.9$\pm$0.1\\
03/08/2022 & 19:21:00 & NOEMA &101.75 &0.6$\pm$0.1\\
20/08/2022 & 11:09:00 & NOEMA &86.25 &0.51$\pm$0.07\\
20/08/2022 & 11:09:00 & NOEMA &101.75 &0.36$\pm$0.06\\
23/08/2022 & 16:54:47 & AMI-LA &15.5 &1.12$\pm$0.09 \\
27/08/2022 & 13:28:34 & AMI-LA &15.5 &1.08$\pm$0.09 \\
01/09/2022 & 13:14:55 & AMI-LA &15.5 &1.19$\pm$0.09 \\
08/09/2022 & 11:58:30 & AMI-LA &15.5 &0.98$\pm$0.08 \\
10/09/2022 & 10:58:47 & AMI-LA &15.5 &1.09$\pm$0.07 \\
10/09/2022 & 15:03:00 & NOEMA &86.25 &0.43$\pm$0.05\\
10/09/2022 & 15:03:00 & NOEMA &101.75 &0.30$\pm$0.04\\
29/09/2022 & 13:46:00 & NOEMA &86.25 &0.32$\pm$0.05\\
29/09/2022 & 13:46:00 & NOEMA &101.75 &0.28$\pm$0.05\\
07/10/2022 & 10:50:15 & AMI-LA &15.5 &1.18$\pm$0.07 \\
10/10/2022 & 09:59:40 & AMI-LA &15.5 &0.93$\pm$0.07 \\
13/10/2022 & 11:12:38 & AMI-LA &15.5 &1.19$\pm$0.07 \\
15/10/2022 & 07:32:00 & NOEMA &86.25 &0.11$\pm$0.04\\
15/10/2022 & 07:32:00 & NOEMA &101.75 &0.19$\pm$0.05\\
16/10/2022 & 08:52:13 & AMI-LA &15.5 &1.06$\pm$0.06 \\
23/10/2022 & 09:17:32 & AMI-LA &15.5 &1.1$\pm$0.2 \\
26/10/2022 & 13:43:59 & NOEMA &86.25 &0.31$\pm$0.05\\
26/10/2022 & 13:43:59 & NOEMA &101.75 &0.39$\pm$0.06\\
31/10/2022 & 10:54:00 & NOEMA &86.25 &0.25$\pm$0.05\\
31/10/2022 & 10:54:00 & NOEMA &101.75 &0.18$\pm$0.04\\
12/11/2022 & 11:42:17 & AMI-LA &15.5 &1.01$\pm$0.06 \\
14/11/2022 & 08:21:57 & AMI-LA &15.5 &0.97$\pm$0.09 \\
18/11/2022 & 12:08:00 & NOEMA &86.25 &0.18$\pm$0.03\\
18/11/2022 & 12:08:00 & NOEMA &101.75 & $<$0.3\\
21/11/2022 & 09:35:09 & AMI-LA &15.5 &1.01$\pm$0.06 \\
01/12/2022 & 07:54:00 & AMI-LA &15.5 &0.94$\pm$0.07 \\
05/12/2022 & 07:20:19 & AMI-LA &15.5 &0.88$\pm$0.07 \\
14/12/2022 & 07:02:53 & AMI-LA &15.5 &$<$1.0 \\
17/12/2022 & 06:33:08 & AMI-LA &15.5 &$<$1.3\\
19/12/2022 & 06:02:28 & AMI-LA &15.5 &1.01$\pm$0.09 \\
27/12/2022 & 06:26:44 & AMI-LA &15.5 &0.96$\pm$0.06 \\
27/12/2022 & 10:11:00 & NOEMA &86.25 &0.14$\pm$0.02\\
27/12/2022 & 10:11:00 & NOEMA &101.75 &0.11$\pm$0.02\\
17/01/2023 & 04:31:15 & AMI-LA &15.5 &0.89$\pm$0.06 \\
20/01/2023 & 03:52:33 & AMI-LA &15.5 &0.98$\pm$0.07 \\
05/02/2023 & 02:59:36 & AMI-LA &15.5 &0.80$\pm$0.05 \\
09/02/2023 & 02:53:50 & AMI-LA &15.5 &0.91$\pm$0.06 \\
12/02/2023 & 02:49:02 & AMI-LA &15.5 &0.84$\pm$0.06 \\
16/02/2023 & 00:04:00 & NOEMA &86.25 &0.11$\pm$0.01\\
16/02/2023 & 00:04:00 & NOEMA &101.75 &0.07$\pm$0.01\\
17/02/2023 & 23:42:56 & MeerKAT &1.28 &0.09$\pm$0.01\\
04/03/2023 & 01:30:23 & AMI-LA &15.5 &0.66$\pm$0.04 \\
10/03/2023 & 01:02:48 & AMI-LA &15.5 &0.5$\pm$0.1 \\
28/03/2023 & 00:25:57 & AMI-LA &15.5 &0.64$\pm$0.04 \\
06/04/2023 & 23:16:42 & AMI-LA &15.5 &0.52$\pm$0.05 \\
07/04/2023 & 23:12:47 & AMI-LA &15.5 &0.69$\pm$0.05 \\
15/04/2023 & 22:15:23 & AMI-LA &15.5 &0.50$\pm$0.05 \\
17/04/2023 & 23:45:15 & AMI-LA &15.5 &0.42$\pm$0.05 \\
21/04/2023 & 23:58:28 & AMI-LA &15.5 &0.58$\pm$0.04 \\
23/04/2023 & 21:58:00 & NOEMA &86.25 &0.08$\pm$0.02\\
23/04/2023 & 21:58:00 & NOEMA &101.75 &0.06$\pm$0.02\\
24/04/2023 & 23:36:11 & AMI-LA &15.5 &0.52$\pm$0.04 \\
05/05/2023 & 20:31:20 & AMI-LA &15.5 &0.38$\pm$0.05 \\
08/05/2023 & 19:37:39 & AMI-LA &15.5 &0.51$\pm$0.08 \\
10/05/2023 & 01:34:14 & AMI-LA &15.5 &0.43$\pm$0.04 \\
11/05/2023 & 20:33:10 & AMI-LA &15.5 &0.55$\pm$0.06 \\
14/05/2023 & 21:47:08 & AMI-LA &15.5 &0.45$\pm$0.04 \\
01/06/2023 & 17:36:52 & AMI-LA &15.5 &0.46$\pm$0.04 \\
04/06/2023 & 18:45:50 & AMI-LA &15.5 &0.34$\pm$0.07 \\
10/06/2023 & 18:46:11 & AMI-LA &15.5 &$<$0.3 \\
08/07/2023 & 16:45:07 & AMI-LA &15.5 &0.23$\pm$0.07 \\
10/07/2023 & 17:23:08 & AMI-LA &15.5 &0.39$\pm$0.08 \\
21/07/2023 & 15:18:06 & AMI-LA &15.5 &0.26$\pm$0.04 \\
12/08/2023 & 16:22:12 & AMI-LA &15.5 &0.34$\pm$0.07 \\
26/08/2023 & 14:52:14 & AMI-LA &15.5 &$<$0.3 \\
02/09/2023 & 18:02:07 & AMI-LA &15.5 &0.30$\pm$0.08 \\
16/09/2023 & 12:46:47 & AMI-LA &15.5 &0.20$\pm$0.05 \\
23/09/2023 & 13:39:13 & AMI-LA &15.5 &0.26$\pm$0.05 \\
28/09/2023 & 13:13:36 & MeerKAT &1.28 &0.10$\pm$0.01\\
29/09/2023 & 16:43:23 & AMI-LA &15.5 &0.27$\pm$0.07 \\
03/10/2023 & 04:10:00 & \textit{e}-MERLIN & 5 &0.35$\pm$0.02\\
06/10/2023 & 09:17:48 & MeerKAT & 3 &0.17$\pm$0.02\\
07/10/2023 & 11:32:12 & AMI-LA &15.5 &0.22$\pm$0.07 \\
15/10/2023 & 10:15:52 & AMI-LA &15.5 &0.24$\pm$0.04 \\
21/10/2023 & 11:36:00 & AMI-LA &15.5 &$<$0.9 \\
28/10/2023 & 09:24:45 & AMI-LA &15.5 &$<$0.15 \\
06/11/2023 & 11:57:51 & AMI-LA &15.5 &0.21$\pm$0.04 \\
12/11/2023 & 11:37:156 & AMI-LA &15.5 &0.33$\pm$0.06 \\
18/11/2023 & 11:00:41 & AMI-LA &15.5 &$<$0.3 \\
18/11/2023 & 01:50:00 & \textit{e}-MERLIN &5 &0.43$\pm$0.03\\
26/11/2023 & 09:44:22 & AMI-LA &15.5 &0.23$\pm$0.04 \\
18/12/2023 & 10:10:33 & AMI-LA &15.5 &0.17$\pm$0.04 \\
07/01/2024 & 08:54:54 & AMI-LA &15.5 &0.13$\pm$0.03 \\
14/01/2024 & 09:14:15 & AMI-LA &15.5 &0.17$\pm$0.04 \\
24/01/2024 & 15:30:00 & \textit{e}-MERLIN &5 &0.31$\pm$0.02\\
28/01/2024 & 03:17:02 & AMI-LA &15.5 &0.21$\pm$0.05 \\
02/02/2024 & 02:37:38 & MeerKAT &1.28 &$<$0.09\\
04/02/2024 & 01:52:43 & MeerKAT &3 &0.30$\pm$0.04\\
31/03/2024 & 00:03:20 & \textit{e}-MERLIN &5 &0.28$\pm$0.02\\
01/06/2024 & 11:05:00 & \textit{e}-MERLIN &5 &0.25$\pm$0.02\\
28/06/2024 & 16:41:25 & MeerKAT &1.28 &0.11$\pm$0.01\\
28/06/2024 & 18:11:31 & MeerKAT &3 &0.39$\pm$0.04\\
08/08/2024 & 11:05:00 & \textit{e}-MERLIN &5 &0.22$\pm$0.02\\
10/08/2024 & 13:16:30 & MeerKAT &1.28 &0.09$\pm$0.02\\
10/08/2024 & 14:47:34 & MeerKAT &3 &0.33$\pm$0.04\\
05/10/2024 & 09:26:25 & MeerKAT &1.28 &0.13$\pm$0.02\\
05/10/2024 & 10:59:27 & MeerKAT &3 &0.22$\pm$0.03\\
14/12/2024$^{a}$ & 08:00:00 & AMI--LA & 15.5 & 0.10$\pm$0.03\\
01/08/2024$^{b}$ & 06:15:00 & AMI--LA  & 15.5 & 0.10$\pm$0.03\\
11/04/2025$^{c}$ & 22:15:00 & AMI--LA  & 15.5 & 0.11$\pm$0.03\\

\end{longtable}
\end{center}

\bibliography{sample631}{}
\bibliographystyle{aasjournal}



\end{document}